\documentclass[sn-mathphys-num]{sn-jnl}

\usepackage{graphicx}%
\usepackage{multirow}%
\usepackage{amsmath, amssymb, amsfonts}%
\usepackage{amsthm}%
\usepackage{mathrsfs}%
\usepackage[title]{appendix}%
\usepackage{xcolor}%
\usepackage{textcomp}%
\usepackage{manyfoot}%
\usepackage{booktabs}%
\usepackage{algorithm}%
\usepackage{algorithmicx}%
\usepackage{algpseudocode}%
\usepackage{listings}%
\usepackage{footnote}
\usepackage{listings}
\usepackage{adjustbox}
\usepackage{amsmath}
\usepackage{bm}

\theoremstyle{thmstyleone}%
\theoremstyle{thmstyletwo}%

\theoremstyle{thmstylethree}%

\begin{document}

\title{FluxMC: Rapid and High-Fidelity Inference for Space-Based Gravitational-Wave Observations}

\author[1,2,3]{\fnm{Bo} \sur{Liang}}

\author[4]{\fnm{Chang} \sur{Liu}}

\author[9]{\fnm{Hanlin} \sur{Song}}

\author[2]{\fnm{Tianyu}  \sur{Zhao}}

\author[2]{\fnm{Minghui} \sur{Du}}
\email{duminghui@imech.ac.cn}

\author[3]{\fnm{He} \sur{Wang}}


\author[8]{\fnm{Haohao} \sur{Gu}}

\author[8]{\fnm{Sensen} \sur{He}}

\author[2,3]{\fnm{Yuxiang} \sur{Xu}}

\author[5]{\fnm{Wei-Liang} \sur{Qian}}

\author[4]{\fnm{Li-e} \sur{Qiang}}

\author[2,3,6,7]{\fnm{Peng} \sur{Xu}}
\email{xupeng@imech.ac.cn}

\author[2,3,6]{\fnm{Ziren} \sur{Luo}}

\author[1]{\fnm{Mingming} \sur{Sun}}
\email{sunmingming@bimsa.cn}


\affil[1]{\orgname{Beijing Institute of Mathematical Sciences and Applications}, \orgaddress{\street{No. 544, Hefangkou Village, Huaibei Town, Huairou District}, \city{Beijing}, \postcode{101408}, \country{China}}}


\affil[2]{\orgdiv{Center for Gravitational Wave Experiment, National Microgravity Laboratory, Institute of Mechanics}, \orgname{Chinese Academy of Sciences}, \orgaddress{\city{Beijing}, \postcode{100190}, \country{China}}}

\affil[3]{\orgdiv{Taiji Laboratory for Gravitational Wave Universe (Beijing/Hangzhou)}, \orgname{University of Chinese Academy of Sciences (UCAS)}, \orgaddress{\city{Beijing}, \postcode{100049}, \country{China}}}


\affil[4]{\orgdiv{National Space Science Center}, \orgname{Chinese Academy
of Sciences}, \orgaddress{\city{Beijing}, \postcode{100190},  \country{China}}}

\affil[5]{\orgdiv{Escola de Engenharia de Lorena}, \orgname{Universidade de São Paulo}, \orgaddress{\city{Lorena}, \postcode{12602-810},
\state{SP}, \country{Brazil}}}

\affil[6]{\orgdiv{Key Laboratory of Gravitational Wave Precision Measurement of Zhejiang Province}, \orgname{Hangzhou Institute for Advanced Study, UCAS}, \orgaddress{\city{Hangzhou}, \postcode{310024},  \country{China}}}

\affil[7]{\orgdiv{Lanzhou Center of Theoretical Physics}, \orgname{Lanzhou University}, \orgaddress{\city{Lanzhou}, \postcode{730000},  \country{China}}}

\affil[8]{\orgdiv{Baidu Inc.}, \orgname{Baidu}, \orgaddress{\city{Beijing}, \postcode{100085}, \country{China}}}

\affil[9]{\orgdiv{School of Physics}, \orgname{Peking University}, \orgaddress{\city{Beijing}, \postcode{100871}, \country{China}}}


\abstract{
Bayesian inference in the physical sciences faces a fundamental challenge: the imperative for high-fidelity physical modeling often clashes with the intrinsic limitations of stochastic sampling algorithms. 
Complex, high-dimensional parameter spaces expose the universal vulnerability of conventional methods, \emph{e.g.}, Markov Chain Monte Carlo (MCMC), which struggle with the prohibitive costs of likelihood evaluations and the risk of entrapment in local optima. 
To resolve this impasse, we introduce \textbf{FluxMC} (\textbf{Fl}ow-guided \textbf{U}nbiased e\textbf{X}ploration \textbf{M}onte \textbf{C}arlo), a machine learning-enhanced framework designed to shift the inference paradigm from blind local search to globally guided transport.
It integrates Flow Matching with Parallel Tempering MCMC, effectively combining the global foresight of generative AI with the rigorous asymptotic convergence and local robustness of temperature-based sampling.
We showcase the efficacy of this framework through the lens of space-based gravitational-wave (GW) astronomy—a field representing the frontier of challenging parameter inversion.
In the analysis of massive black hole binaries using high-fidelity waveforms (\texttt{IMRPhenomHM}), FluxMC achieves robust convergence in under five hours, whereas traditional Parallel Tempering MCMC fails to converge even after hundreds of hours, yielding high Jensen-Shannon divergences (JSD) of $\mathcal{O}(10^{-1})$. 
Our method reduces the distributional error by two to three orders of magnitude.
Furthermore, for computationally efficient models (\texttt{IMRPhenomD}), it eliminates systematic biases caused by local-optima entrapment. 
Ultimately, FluxMC removes the necessity to compromise between model accuracy and analysis speed, establishing a new computational foundation where scientific discovery is limited only by observational data quality, not by algorithmic capacity.
}




\maketitle

\section*{Introduction}

\section{Introduction}


Bayesian inference stands as the mathematical cornerstone for data-driven discovery across the natural sciences—from reconstructing high-resolution medical images~\cite{sd_sw} and characterizing complex quantum states~\cite{Arrazola_2019, Layden2022QuantumenhancedMC} to mapping the universe~\cite{Lewis:2002ah, Loredo1992:promise}. It provides a rigorous probabilistic framework for extracting model parameters from noisy observational data.
Remarkably, it has played a critical role in interpreting the  hundreds of gravitational wave (GW) signals captured  by the LIGO-Virgo-KAGRA ground-based GW observatories~\cite{Thrane2019:an,Dong:2025igh}. 
The  posterior distributions of parameters are typically acquired  via stochastic sampling algorithms, most notably Markov Chain Monte Carlo (MCMC)~\cite{Metropolis1953:equation,Hastings1970:monte}.
However, this methodology faces  profound challenges in two scenarios: firstly,  when the sampling becomes prohibitively expensive due to computationally intensive models with high-dimensional parameters (\emph{e.g.} $> 10$ dimensions); and secondly,  when the posterior exhibits a complex landscape featuring degeneracy or multimodality (\emph{i.e.}  multiple separated high-probability regions). 
For  multimodal posteriors, standard MCMC struggles to  discover all modes~\cite{Zhou2011:multi} and to  move effectively  among them~\cite{Dawn2009:sufficient}. 
Collectively, these issues can  lead to  trapping to local optima, biased estimation,  slow convergence, and consequently failure to characterize the full target distribution~\cite{Arai:2025quantum}.


\begin{figure}[t!]
    \centering
    \includegraphics[width=0.95\textwidth]{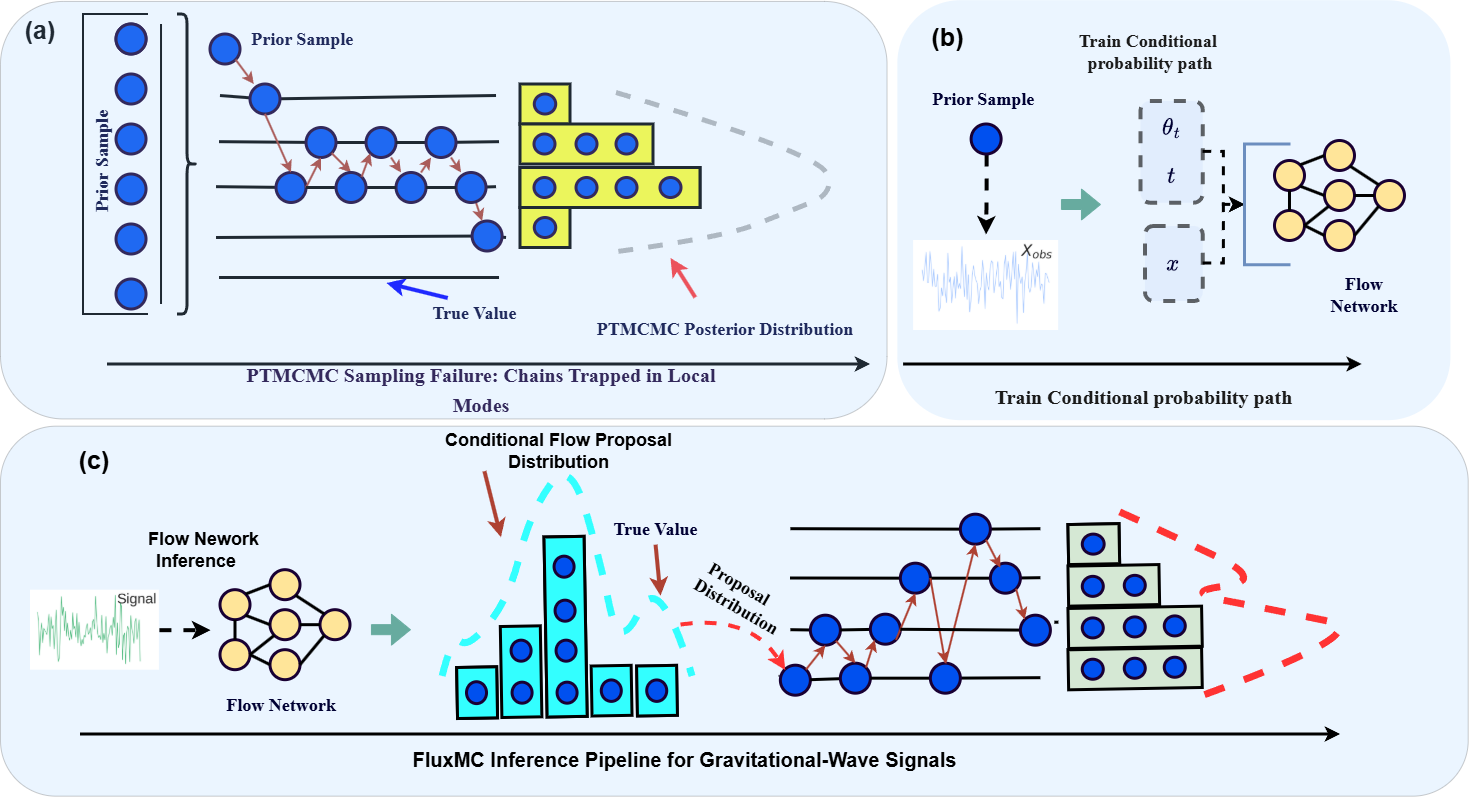}
    \caption{
    \textbf{Conceptual overview of the FluxMC framework enabling rapid, unbiased global inference.}
    (a) \textbf{The Challenge}: The Multimodal Trap. Traditional sampling methods (e.g., PTMCMC) act as ``blind walkers.'' They frequently get trapped in local optima of complex, high-dimensional landscapes, failing to bridge low-probability gaps. This results in systematically biased conclusions and prohibitively slow convergence (weeks/months).
    (b) \textbf{Conditional Generative Modeling}: We formulate sampling as a conditional generative task. Conditioned on observed data $x_{\text{obs}}$, a FM network learns the continuous probability paths (vector fields) that transform a simple prior distribution into the complex, multi-modal posterior distribution.
    (c) \textbf{The Solution}: Flow-Guided Global Proposals. During inference, the trained flow network acts as a ``smart'' proposal distribution, allowing the MCMC sampler (FluxMC) to efficiently jump between isolated modes. This overcomes sampling barriers, achieving accurate, unbiased recovery of the true value with rapid convergence.
    }
    \label{fig:model}
\end{figure}

These  challenges are particularly pronounced in the era of space-based GW detecton. 
Following the breakthrough detections by ground-based interferometers~\cite{Aasi_2015, Acernese_2015, 10.1093/ptep/ptaa125}, 
the field is poised to enter a new regime with future space-based observatories such as the Laser Interferometer Space Antenna (LISA)~\cite{amaro2017laser,baker2019laser}, Taiji~\cite{hu2017taiji,luo2021taiji,doi:10.34133/research.1252}, and TianQin~\cite{luo2016tianqin} in the next decade. 
These missions will unlock the millihertz frequency band, enabling  the detection of massive black hole binary (MBHB) mergers~\cite{klein2016science}.
As the most energetic astrophysical events since the Big Bang, MBHB mergers can serve as a probe to discriminate between different hierarchical formation models of massive black holes~\cite{shen2026revealing},  and may serve as standard sirens to help  resolve the ``Hubble tension'' problem~\cite{Kyutoku:2016zxn}. 
Specifically, the precise measurement of their masses and spins is essential  for discriminating between various  black hole seed mechanisms,  
and for performing inspiral-merger-ringdown consistency tests to probe deviations from general relativity~\cite{Colpi2024:lisa,LISA:2022yao}.
Moreover, as standard sirens, coalescing MBHBs enable direct determination of luminosity distances, provided their  masses can be accurately recovered~\cite{Schutz1986:determining}.
Consequently, the robustness of  sampling method becomes the key factor in capturing source properties and  fulfilling the mission's scientific objectives.   
However, realizing this potential faces the well‑known multimodality in  MBHB posteriors~\cite{marsat2021exploring}, and the high computational cost of high-fidelity waveform models that include higher‑order modes (HMs)~\cite{Kalaghatgi:2019log,PhysRevLett.120.161102,Pompili:2023tna}. 
With an anticipated detection rate of $\mathcal{O}(10^2)$ events per year and signal durations spanning months, the cumulative cost of data analysis is prohibitive.

Currently, the stat-of-the-art method for exploring these complex  posteriors is Parallel Tempering Markov Chain Monte Carlo (PTMCMC)~\cite{justin_ellis_2017_1037579}. 
By running multiple chains at different ``temperatures'' to flatten the likelihood surface, PTMCMC dominates the analysis pipelines of modern astronomy due to its robustness in traversing multi-modal distributions. 
However, PTMCMC acts as a ``blind'' explorer; it relies on local random walks that scale poorly with parameter space dimensionality and complexity. 
When applied to high-fidelity waveform models that incorporate HMs, PTMCMC requires hundreds of hours to analyze a single event. 
This latency renders real-time analysis impossible and creates an unmanageable backlog for mission catalog generation. 
Furthermore, the inherent limitations of PTMCMC in navigating pathological landscapes manifest either as severe distortions in specific extrinsic parameters—most notably the coalescence phase $\phi_c$ and polarization $\psi$—or as a failure to achieve reliable convergence within reasonable timescales when high-fidelity HM templates are utilized.

To overcome these fundamental limitations, we introduce \textbf{FluxMC} (\textbf{Fl}ow-guided \textbf{U}nbiased e\textbf{X}ploration \textbf{M}onte \textbf{C}arlo). 
Recently, similar approaches that couple MCMC with normalizing flows have shown great promise in accelerating parameter estimation for ground-based gravitational-wave detectors~\cite{wong2022flowmcnormalizingflowenhancedsampling, wong2023fastgravitationalwaveparameter, wouters2025robustparameterestimationminutes, doi:10.1073/pnas.2109420119}. 
Building upon this powerful generative paradigm, our framework extends the frontier to the unique and highly degenerate multimodal sampling challenges of space-based observatories.
As conceptually illustrated on a Gaussian mixture benchmark (see Fig.~\ref{fig:gmm}), our framework addresses the multimodal sampling challenge by augmenting the robustness of Parallel Tempering with the global foresight of deep generative modeling. 
While traditional PTMCMC chains often become trapped in local optima (Fig.~\ref{fig:model}a), FluxMC utilizes a Flow Matching (FM) network~\cite{liu2023flow,lipman2023flow} trained on the specific observation  to learn the global structure of degeneracies within the posterior (Fig.~\ref{fig:model}b). 
By employing this trained flow as a ``smart'' global proposal mechanism within the MCMC sampler, FluxMC enables efficient jumps between isolated modes, bypassing the local barriers that trap traditional algorithms (Fig.~\ref{fig:model}c). 

We demonstrate that FluxMC effectively navigates these trade-offs across multiple regimes. Using the computationally efficient IMRPhenomD model, we show that FluxMC resolves critical sampling failures where traditional PTMCMC becomes trapped in local optima of extrinsic parameters, ensuring unbiased recovery of the global truth. Moving to high-fidelity IMRPhenomHM signals—where standard methods are prohibitively slow—FluxMC achieves robust convergence in under five hours, reducing the distributional error (JS divergence) by a factor of $\sim 180$ compared to non-converged PTMCMC results (i.e., yielding distorted distributions when exploring the full parameter space). This framework reduces inference time from weeks to merely a few hours, satisfying the latency requirements for real-time multi-messenger alerts.
FluxMC is designed as a general-purpose algorithm, offering a pathway to rapid, unbiased global inference for the broader scientific community facing complex inverse problems.

The remainder of this paper is organized as follows. We first validate FluxMC on a Gaussian mixture benchmark to demonstrate its fundamental ability to overcome multimodal traps. We then present detailed case studies for both LISA and Taiji missions using the aforementioned waveform templates to assess robustness and speedup. The complete methodological framework, including the flow-guided proposal strategy and adaptive training, is detailed in the Methods section.

\section{Results}\label{sec2}

\subsection{Overcoming Multimodal Traps with FluxMC: A Gaussian  Mixture Benchmark Demonstration}

FluxMC is designed to bridge the gap between rigorous Bayesian sampling and modern generative AI, effectively combining the ``best of both worlds.''
Functionally, it inherits the theoretical guarantees of Parallel Tempering MCMC (PTMCMC), maintaining the property of asymptotic exactness for unbiased recovery of the true posterior distribution, while leveraging temperature ladders to flatten complex likelihood surfaces and prevent chains from becoming stuck in local modes.

However, traditional PTMCMC remains fundamentally limited by its ``blind'' local proposal mechanism, which struggles to cross vast low-probability barriers in pathological landscapes. FluxMC revolutionizes this paradigm by integrating FM as a ``Global Navigator''. Instead of relying on random local walks, FluxMC utilizes the learned global structure of degeneracies to generate smart, non-local proposals. This integration introduces two transformative advantages: global exploration capabilities and orders-of-magnitude acceleration. The flow-guided proposals allow samplers to ``tunnel'' through probability barriers, enabling immediate transitions between isolated high-probability modes that remain inaccessible to standard algorithms. Simultaneously, by bypassing inefficient burn-in phases, FluxMC reduces convergence time from weeks to minutes, satisfying the critical latency requirements for real-time multi-messenger astronomy.
\begin{figure}[htb!]
    \centering
    \includegraphics[width=0.95\textwidth]{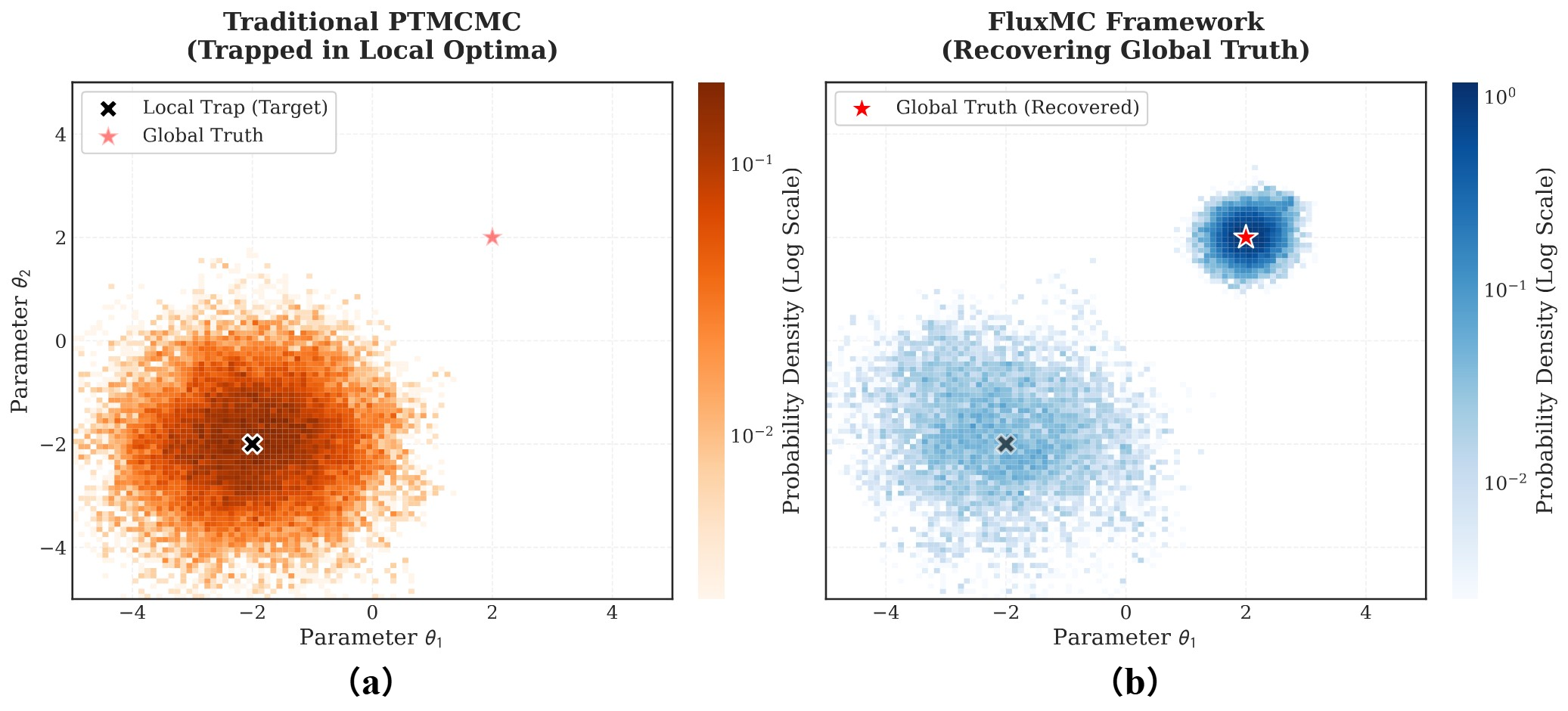}
    \caption{
    \textbf{The Gaussian Mixture Benchmark.}
    To test robustness, all chains are initialized within the suboptimal \textbf{Local Trap}.
    \textbf{(a) Traditional PTMCMC Failure.} Confined by the initialization, the sampler is trapped in the local basin (orange). Despite parallel tempering, it fails to tunnel through the probability barrier.
    \textbf{(b) FluxMC Success.} FluxMC (blue) leverages FM to "preview" the global structure. The flow-guided proposals allow the sampler to escape the initialization trap and recover the Global Truth immediately.
    }
    \label{fig:gmm}
\end{figure}
To rigorously demonstrate these capabilities in severe multi-modal regimes, we constructed a Gaussian Mixture Benchmark designed to mimic the challenges of GW degeneracies~\cite{marsat2018fourier}.
The target probability density $p(\mathbf{x})$ is defined as an equal-weighted mixture of two isotropic Gaussians:
\begin{equation}
    p(\mathbf{x}) \propto \mathcal{N}(\mathbf{x}; \boldsymbol{\mu}_{\text{trap}}, \sigma_{\text{trap}}^2 \mathbf{I}) + \mathcal{N}(\mathbf{x}; \boldsymbol{\mu}_{\text{truth}}, \sigma_{\text{truth}}^2 \mathbf{I})
\end{equation}
where the landscape features a broad, suboptimal Local Trap ($\boldsymbol{\mu}_{\text{trap}}=[-2, -2], \sigma_{\text{trap}}=1.0$) and a sharp, high-probability Global Truth ($\boldsymbol{\mu}_{\text{truth}}=[2, 2], \sigma_{\text{truth}}=0.3$), separated by a significant probability barrier.
Crucially, to simulate a ``worst-case'' initialization scenario, we initiate all sampling chains deep within the basin of the Local Trap.

As illustrated in Fig.~\ref{fig:gmm}A, traditional PTMCMC exhibits  insufficient sampling capabilities under this initialization. 
PTMCMC sampler explores only the vicinity of its starting point. 
It remains trapped in the broader local basin, failing to cross the barrier to the true mode.
This results in an \textbf{overconfidently biased} posterior that completely misses the true parameters.

In contrast, FluxMC faithfully recovers both modes (Fig.~\ref{fig:gmm}B). This success stems from the method's ability to enhance the search using generative AI. By employing FM~\cite{lipman2023flow,tong2023improving}, the model learns a  vector field that captures the approximate global density before sampling begins. Consequently, the flow-generated proposals act as non-local jumps, allowing the sampler to ignore the local barrier and immediately access the high-probability region, effectively preventing entrapment in the initialization basin.

\subsection{Improving Sampling Efficiency and Robustness for Efficient Waveforms}
\begin{figure}[h!]
    \centering
    \includegraphics[width=0.8\textwidth]{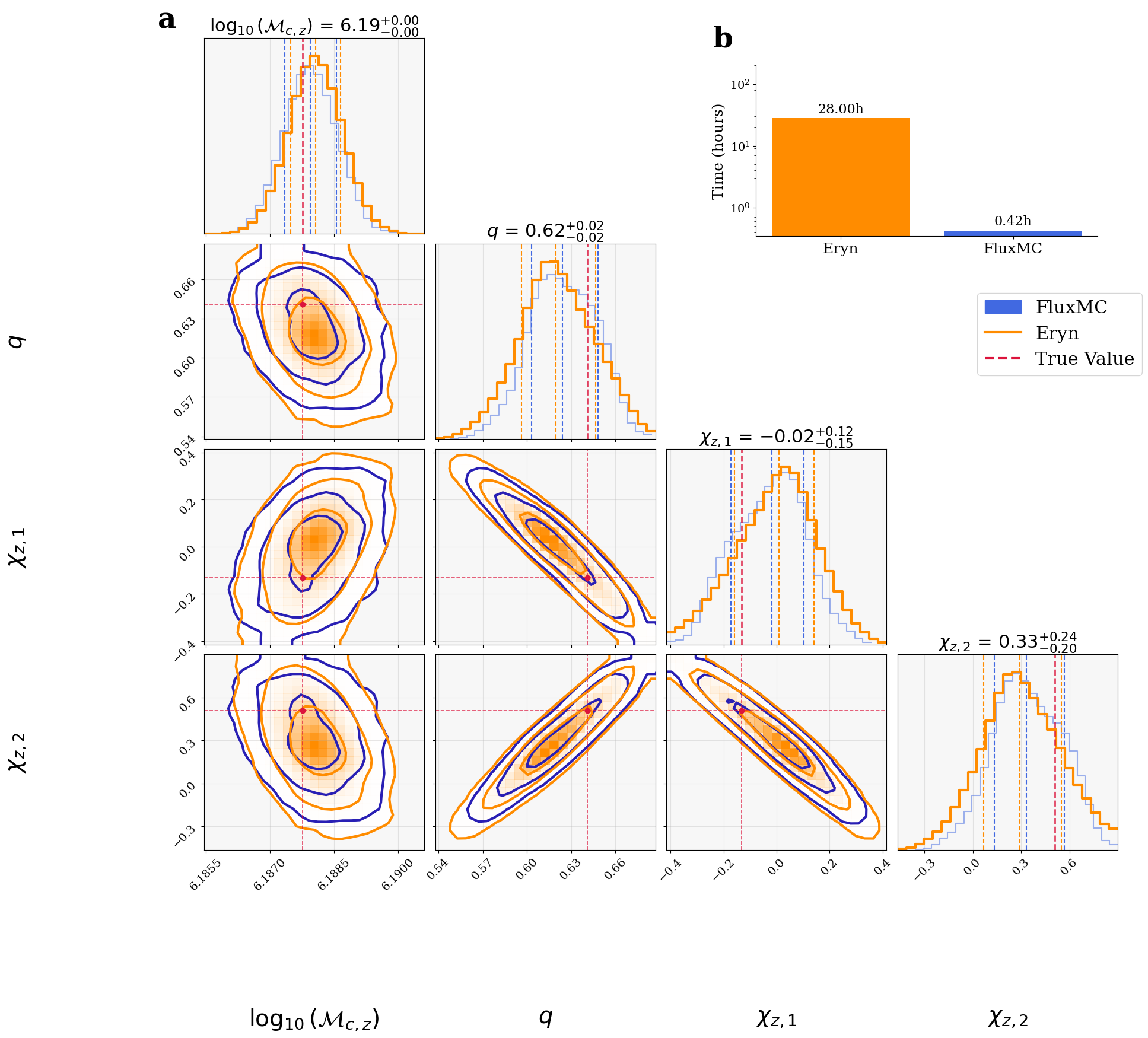} 
    \caption{
        \textbf{FluxMC Improving Sampling Efficiency for Efficient Waveforms.}
        Comparison of parameter estimation for a simulated MBHB signal, using the computationally efficient \texttt{IMRPhenomD} template.
        \textbf{(a)} Corner plot showing 1D and 2D posterior distributions. Both FluxMC (blue) and PTMCMC (orange) recover the true injected parameters (red dashed lines).
        \textbf{(b)} Comparison of total computational time. \textbf{FluxMC} ($\sim$0.42 hours) is $\sim$67 times faster than PTMCMC (28.00 hours), demonstrating extreme efficiency gains in standard inference scenarios.
    }
    \label{fig:lisa_unbiased_d_model}
\end{figure}
While HMs present the ultimate computational challenge, the difficulties of Bayesian inference persist even when employing computationally efficient templates like \texttt{IMRPhenomD}~\cite{khan2016frequency}. In this section, we demonstrate that \textbf{FluxMC} not only dramatically accelerates parameter estimation for these standard waveforms but, more importantly, resolves critical sampling failures where traditional methods become trapped in local optima~\cite{karnesis2014bayesian, babak2008mock}.
To validate the universality of our method across different space-borne GW antennas, we utilize orbital parameters and noise models corresponding to representative space-based observatories (e.g., LISA and Taiji) to generate simulated data. Detailed descriptions of the data generation, detector response, and preprocessing steps are provided in the Methods section~\ref{Sec:model}.

\subsubsection{Accelerating Inference without Compromising Accuracy}

One of \textbf{FluxMC}'s primary advantages is its ability to drastically reduce the computational cost of inference. We first consider a ``best-case'' scenario for traditional methods: a simulated MBHB event where the posterior distribution is relatively unimodal and well-behaved.

As shown in Fig.~\ref{fig:lisa_unbiased_d_model}a, both \textbf{FluxMC} (blue) and the conventional PTMCMC sampler (orange) successfully recover the true injected parameters (red dashed lines). The posteriors for key parameters, including the redshifted chirp mass $\log_{10}(\mathcal{M}_{c,z})$, mass ratio $q$, and spin components $\chi_{z,1}, \chi_{z,2}$, are consistent between the two methods and well-centered on the true values. 

However, a stark contrast emerges in the computational efficiency. As illustrated in Fig.~\ref{fig:lisa_unbiased_d_model}b, while the PTMCMC sampler requires \textbf{28.00 hours} to achieve convergence, \textbf{FluxMC} produces the same high-fidelity result in just \textbf{0.42 hours} ($\sim$25 minutes). This represents a remarkable \textbf{$\sim$67-fold reduction} in computational cost. This result confirms that in scenarios where traditional sampling is successful, \textbf{FluxMC} provides a significant speedup without any loss of accuracy. 
To rigorously validate the statistical consistency and unbiasedness of our framework across the parameter space, we performed an injection campaign using 30 simulated MBHB signals. The posterior calibration (P-P) test demonstrates that FluxMC produces perfectly calibrated posteriors, whereas traditional PTMCMC exhibits statistically significant miscalibration (see Appendix~\ref{sec:posterior_calibration} and Fig.~\ref{fig:pp_plot} for details).
The complete posterior distributions for this event are detailed in Appendix Fig.~\ref{fig:lisa_unbiased_d_full}.

\subsubsection{Overcoming Sampling Failures in Complex Multimodal Posteriors}

\begin{figure}[h!]
    \centering
    \includegraphics[width=1.0\textwidth]{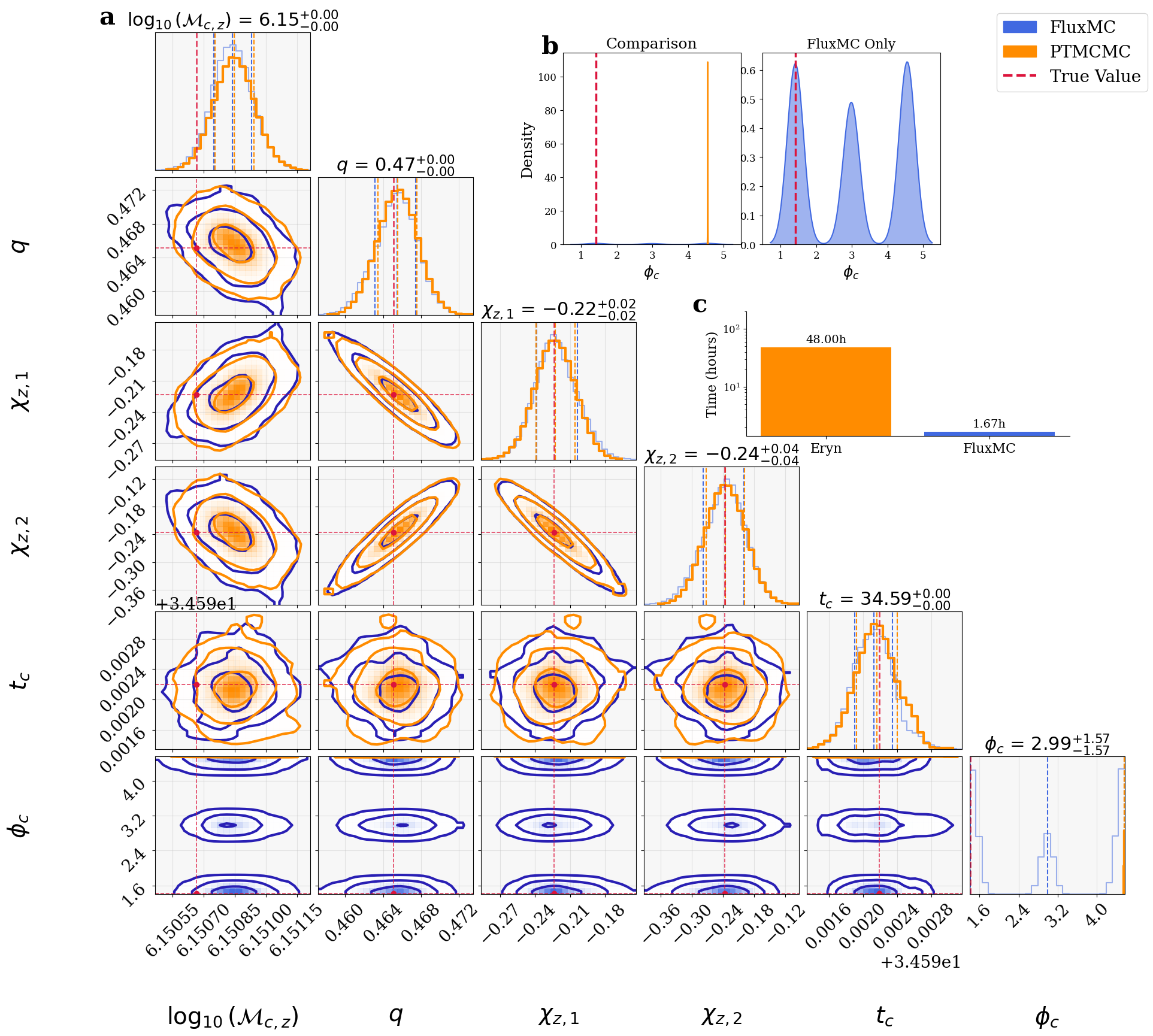} 
    \caption{
        \textbf{Robust Inference in Simulation Case \textrm{I}.}
        \textbf{(a)} Corner plot showing PTMCMC (orange) converging to a local optimum, missing the true values (red dashed lines) for parameters such as $\phi_c$.
        FluxMC (blue) accurately recovers the true parameters.
        \textbf{(b)} Focused view of the coalescence phase ($\phi_c$), highlighting the severe bias in the PTMCMC result.
        \textbf{(c)} FluxMC (1.7 hours) is over 40 times faster than the biased PTMCMC run (75 hours).
        The complete posterior distributions for all parameters for this event are shown in Appendix Fig.~\ref{fig:lisa_biased_d_full}. 
    }
    \label{fig:lisa_d_model}
\end{figure}
The advantages of FluxMC extend beyond mere acceleration. A more fundamental challenge in GW  inference is the ``sampling failure'' problem, where traditional algorithms fail to navigate the complex landscape of the posterior distribution. 
Even with simplified templates like \texttt{IMRPhenomD}, the posterior distribution for MBHB mergers is rife with global degeneracies—particularly in extrinsic parameters such as sky location and orbital orientation—which create separated modes that trap even global exploration algorithms like PTMCMC.

This failure is strikingly illustrated in a representative space-based simulation shown in Fig.~\ref{fig:lisa_d_model}.
For this event, we injected a signal (see caption for parameters) and performed inference using 5 days of data. When analyzed with the standard PTMCMC sampler (orange), the recovered posterior distributions are confidently biased away from the true values (red dashed lines). This bias is particularly severe for the coalescence phase $\phi_c$ (Fig.~\ref{fig:lisa_d_model}b), a parameter critical for multi-messenger synchronization. The PTMCMC chains have become trapped in a local optimum, resulting in scientifically invalid conclusions.

\begin{figure}[htb!]
    \centering
    \includegraphics[width=1.0\textwidth]{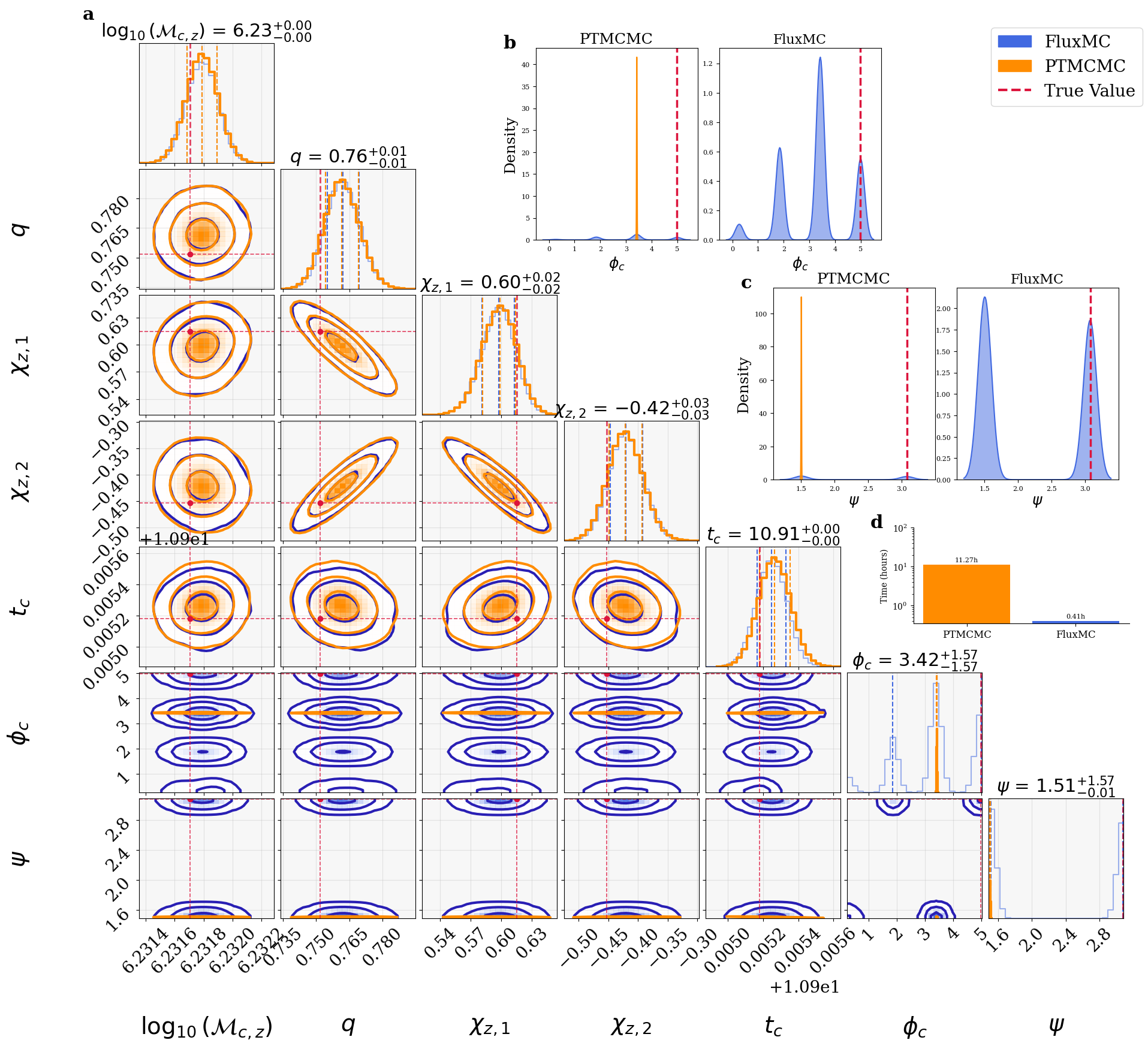}
    \caption{
        \textbf{Robust Inference in Simulation Case \textrm{II}.}
        \textbf{(a)} Corner plot comparing FluxMC (blue) and PTMCMC (orange). 
        \textbf{(b, c)} Focused 1D posteriors for $\phi_c$ and $\psi$, highlighting PTMCMC's inability to explore the multi-modal space.
        \textbf{(d)} Wall-clock time comparison: FluxMC (0.41 hours) is $\sim$27.5 times faster than the biased PTMCMC run (11.27 hours).
        The complete posterior distributions for all parameters for this event are shown in Appendix Fig.~\ref{fig:taiji_unbiased_d_full}. 
    }
    \label{fig:taiji_d_comparison} 
\end{figure}

In contrast, FluxMC (blue) overcomes this barrier using the identical setup. By leveraging the global guidance of the FM model, it correctly identifies the true parameters and produces an unbiased posterior distribution. Furthermore, it achieves this robust result in approximately \textbf{1.7 hours}, compared to the \textbf{75 hours} on the biased PTMCMC run (Fig.~\ref{fig:lisa_d_model}c). This demonstrates that FluxMC is not only faster but fundamentally more robust against the ``local trap'' vulnerability of traditional MCMC.

This limitation of conventional samplers is not restricted to a single instance; rather, it represents a general challenge in Bayesian inference for space-based GW observations. We further demonstrate this pervasive issue in a second, more general example featuring complex multi-modal distributions, where the traditional sampler remains trapped in suboptimal modes.
We demonstrate this in Fig.~\ref{fig:taiji_d_comparison} using a simulated event for Case \textrm{II}. 
Here, the PTMCMC sampler (orange) again fails to locate the global mode. While it finds consistent values for some intrinsic parameters (panel a), it exhibits significant bias in extrinsic parameters such as $\phi_c$ (panel b) and polarization $\psi$ (panel c). FluxMC (blue), however, correctly identifies the multi-modal nature of the posterior, recovering the true parameters within the dominant modes. 

Critically, FluxMC achieves this accurate result in just \textbf{0.41 hours}.
This highlights a fundamental performance gap: while the traditional PTMCMC sampler expended \textbf{11.27 hours} only to yield a scientifically invalid, biased conclusion, FluxMC successfully recovered the global truth with an order-of-magnitude reduction in latency. 
This result underscores that FluxMC provides not just an acceleration, but a necessary capability for robust inference where standard algorithms fail.

These results collectively establish FluxMC as a critical tool for future space-based observatories. By solving the dual challenges of computational cost and sampling bias, it ensures that the expected scientific achievements of future gravitational-wave missions can be fully realized.

\subsection{Enabling Rapid, Robust Inference with High-Fidelity Templates}
The former tests based on \texttt{IMRPhenomD} model have demonstrated  FluxMC's proficiency in sampling multimodal distributions, 
while unlocking the full scientific potential of space-based observatories requires the routine adoption of high-fidelity waveform templates incorporating HMs. 
On the one hand, the contributions of HMs become non-negligible for MBHB signals characterized by high signal-to-noise ratios of up to $\mathcal{O}(10^3)$ - $\mathcal{O}(10^4)$~\cite{Gong:2023ecg}. 
On the other hand, HMs are essential  for breaking the critical degeneracy between luminosity distance ($D_L$) and inclination angle ($\iota$). 
\begin{figure}[h!]
    \centering
    \includegraphics[width=1.0\textwidth]{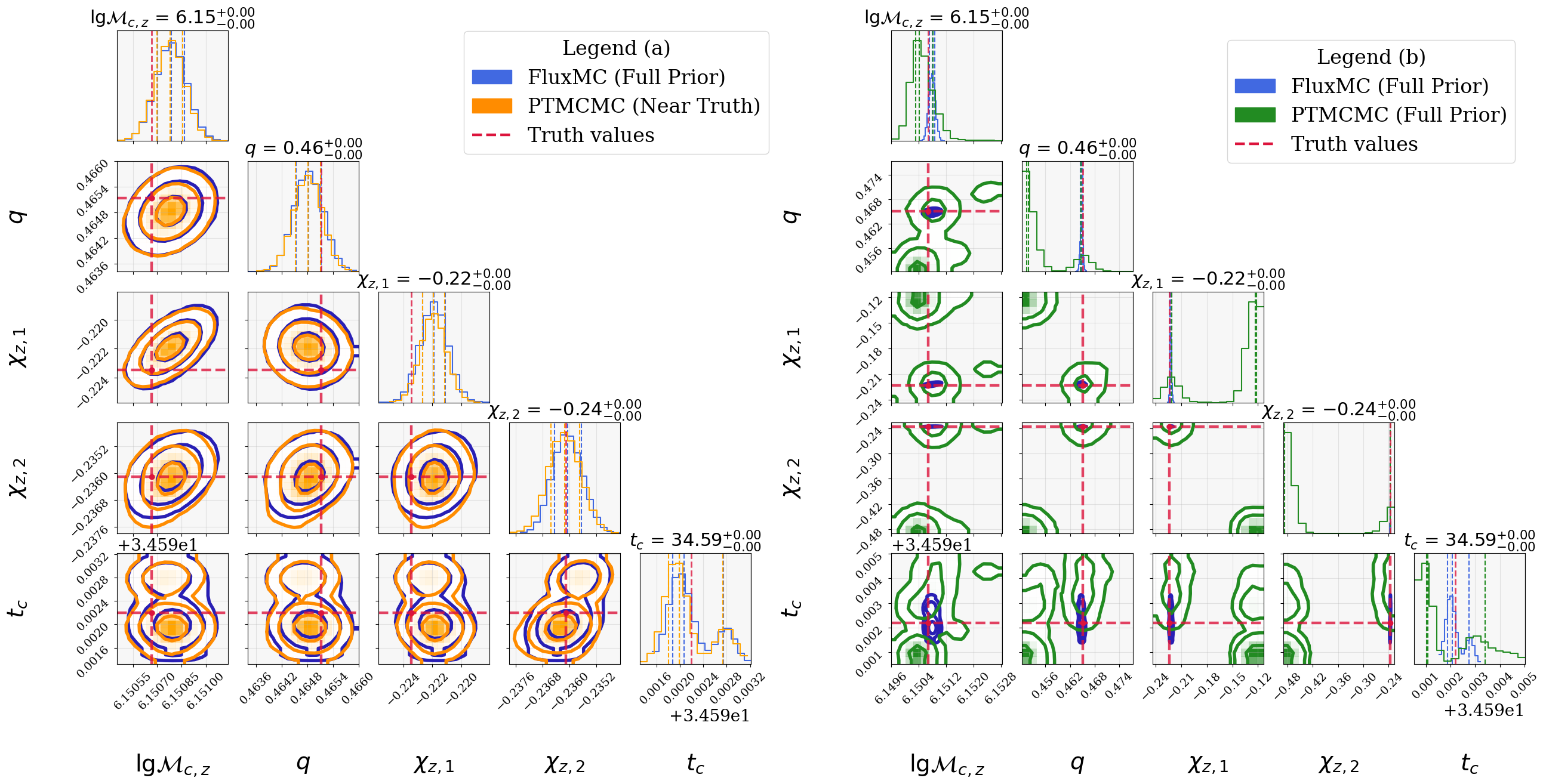} 
    \caption{
    \textbf{FluxMC Enables Rapid and Reliable High-Fidelity \texttt{IMRPhenomHM} Inference (Case \textrm{III}).}
    Comparison of posterior distributions for key intrinsic parameters using the high-fidelity \texttt{IMRPhenomHM} template.
    The true injected parameters (red dashed lines) are: $[\log_{10}(\mathcal{M}_{c,z})=6.150, q=0.465, \chi_{z,1}=-0.223, \chi_{z,2}=-0.236]$.
    \textbf{(a) Validation of Accuracy.} The FluxMC (Full Prior) run (blue) produces distributions statistically consistent with the PTMCMC (Near Truth) benchmark (orange), confirming accurate parameter recovery.
    \textbf{(b) Robustness Comparison.} Despite utilizing an NVIDIA A100 (80GB) GPU for hundreds of hours, the conventional PTMCMC (Full Prior) sampler (green) fails to locate the global optimum. Instead, it converges to a biased solution corresponding to a spurious local mode (visible as offset clusters in $q$ and $\chi_{z}$), completely missing the true parameters. In contrast, FluxMC (blue) successfully recovers the true parameters in a fraction of the time ($\sim$4.5 hours).
    The complete posterior distributions for all parameters for this event are shown in Appendix Fig.~\ref{fig:lisa_hm_validation_full} and Fig.~\ref{fig:lisa_hm_robustness_full}.
    }
    \label{fig:hm_comparison}
\end{figure}
Accurate inference of these parameters is foundational for using MBHBs as ``standard sirens'' to probe the expansion history of the universe. 
Moreover, the constraints on masses and spins are also  significantly enhanced by the inclusion of HMs, allowing for better understanding of massive black hole population models and their hierarchical merger mechanisms~\cite{LISA:2022yao}. 
However, the introduction of HMs significantly increases the complexity of the likelihood surface, creating a multimodal landscape that challenges standard sampling algorithms.

We demonstrate FluxMC's performance in this situation using a high-fidelity MBHB signal for Case \textrm{III}, employing the \texttt{IMRPhenomHM} template. The results are presented in Fig.~\ref{fig:hm_comparison}.

First, to validate accuracy, we compare our method against a ``gold standard'' PTMCMC run initialized tightly around the true values~\cite{Dax_2021, farr2015parameter, marsat2021exploring}. As shown in Fig.~\ref{fig:hm_comparison}a, the FluxMC analysis (blue), which starts from the full unconstrained prior, produces posterior distributions that perfectly overlap with this localized benchmark (orange). This confirms that FluxMC accurately recovers the physical parameters, including the precise mass ratio ($q$) and spin components ($\chi_{z}$), without introducing bias.

Crucially, Fig.~\ref{fig:hm_comparison}b reveals the fundamental limitations of traditional methods in a blind search scenario. Despite being allocated extensive computational resources—an NVIDIA A100 (80GB) GPU running for hundreds of hours—the conventional PTMCMC (Full Prior) sampler (green contours) failed to converge to the true posterior. Instead of exploring the full parameter space, the sampler converged to incorrect local optima. This observation is consistent with previous studies of related LISA-MBHB inference problems~\cite{Marsat:2020rtl, Weaving:2023fji}, which likewise reported multimodality-induced difficulties for conventional samplers.
This is evident in the posterior distributions for mass ratio $q$ and spins $\chi_{z}$, where the PTMCMC results appear as isolated, biased clusters far from the true values (red dashed lines). This behavior indicates that the sampler was unable to jump between the separated modes of the HM likelihood surface, effectively getting stuck in a secondary peak. 
This failure precisely recapitulates the fundamental limitation demonstrated in our Gaussian Mixture Benchmark (Sec.~\ref{fig:gmm}). Lacking a global view of the landscape, the traditional sampler's local stepping mechanism is unable to tunnel through the vast low-probability barriers that separate these degenerate modes, thereby confining the chains to their initialization basin.
In contrast, FluxMC (blue) successfully navigated this multimodal landscape to locate the global optimum in approximately \textbf{4.5 hours}. The complete set of posterior distributions for all 11 parameters is provided in Appendix Fig.~\ref{fig:lisa_hm_validation_full} and Fig.~\ref{fig:lisa_hm_robustness_full}.

\begin{figure}[h!]
    \centering
    \includegraphics[width=1.0\textwidth]{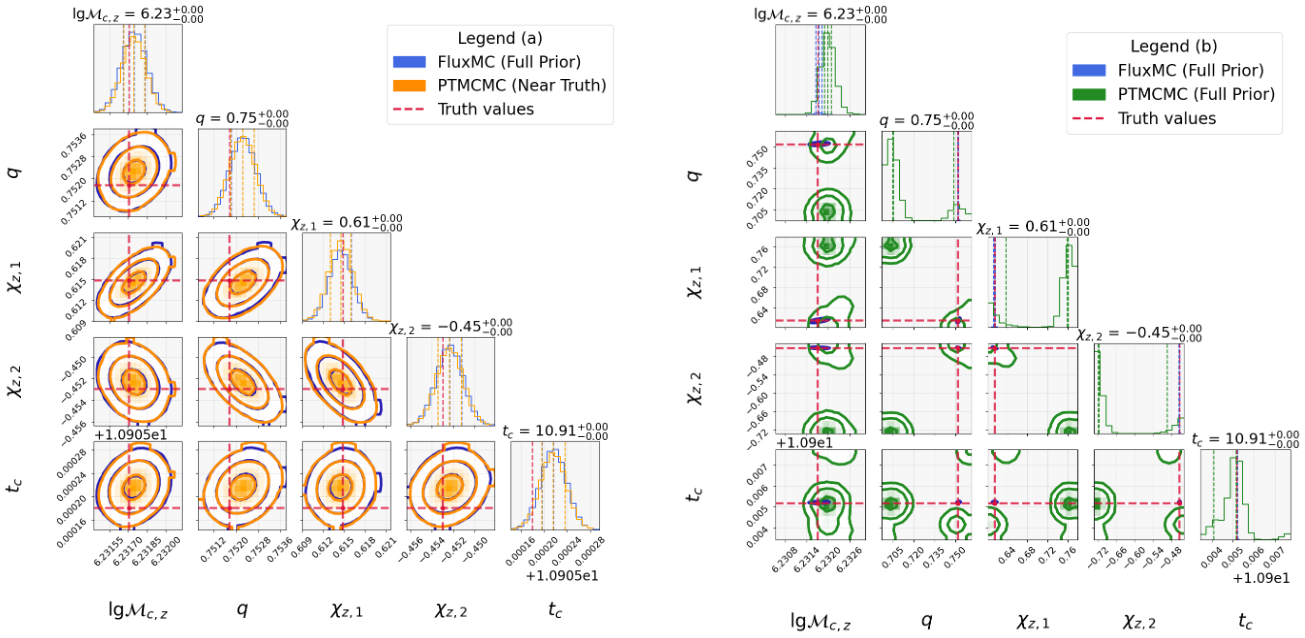} 
\caption{
    \textbf{FluxMC Enables Rapid and Reliable High-Fidelity \texttt{IMRPhenomHM} Inference (Case \textrm{IV}).}
    Replication of the validation for a simulated MBHB signal using \texttt{IMRPhenomHM}.
    \textbf{(a) Validation:} FluxMC (blue) perfectly matches the localized PTMCMC benchmark (orange).
    \textbf{(b) Robustness:} Consistent with the Case \textrm{IV} results, the unconstrained PTMCMC (Full Prior) run (green) fails to break parameter degeneracies. The posterior for $q$ remains multi-modal (bi-modal), oscillating between the truth and a degenerate peak, while spins rail against boundaries. FluxMC (blue) correctly identifies the single global mode and recovers the true parameters.
    The complete posterior distributions for all parameters for this event are shown in Appendix Fig.~\ref{fig:taiji_hm_validation_full} and Fig.~\ref{fig:taiji_hm_robustness_full}.
}
\label{fig:hm_comparison_taiji} 
\end{figure}

This robust performance is not unique to  Case \textrm{III}; 
we observe a similar, decisive advantage in the  analysis for Case \textrm{IV}  (Fig.~\ref{fig:hm_comparison_taiji}). 
Here, the conventional PTMCMC sampler (green) fails to fully exploit the information contained in HMs to break parameter degeneracies. This failure is most evident in the mass ratio $q$, where the PTMCMC posterior remains multi-modal (specifically bi-modal), oscillating between the true solution and a degenerate secondary peak. Furthermore, the spin parameters exhibit significant railing against the physical boundaries. In contrast, FluxMC (blue) effectively breaks this degeneracy, recovering a single, sharp mode centered precisely on the true parameters. The full posterior distributions detailing this comparison are available in Appendix Fig.~\ref{fig:taiji_hm_robustness_full}.

To  quantify the fidelity of the recovered posteriors, we computed the JS divergence between the sampled results and the reference ground-truth distribution for all 11 model parameters. The comprehensive comparison is detailed in Table~\ref{tab:js_comparison_full}. 
For Simulation Case \textrm{III}, conventional PTMCMC exhibits catastrophic divergence across all parameters (mean JSD $\approx 0.726$), particularly for the sky location and distance where divergences exceed $0.8$. 
FluxMC significantly corrects these distributions, achieving a mean JSD of $0.019$.
Empirical validation studies suggest that, particularly for complex, multimodal posterior distributions, a JSD below $0.05$ indicates overall statistical agreement~\cite{PhysRevD.106.104021}.
The improvement is even more pronounced in the analysis for Case \textrm{IV}. 
While PTMCMC fails to break parameter degeneracies (mean JSD $\approx 0.581$), FluxMC achieves near-perfect recovery with an average JSD of only $0.003$—an improvement of approximately $180\times$. 
This JSD of $\mathcal{O}(10^{-3})$ corresponds to a unimodal structure in the extrinsic parameters, demonstrating the potential of higher-order modes to fully break degeneracies.
This confirms that FluxMC's robustness against local optima is consistent across different observatory configurations.

\begin{table*}[t!]
    \centering
    \caption{
    \textbf{Comparison of JSD for Simulation Cases III and IV.} 
    Comparison of JSD between the sampled posteriors (PTMCMC vs. FluxMC) and the reference ground-truth distribution for all 11 inferred parameters. 
    The ``Improv.'' column indicates the reduction factor in divergence ($\text{JSD}_{\text{PT}}/\text{JSD}_{\text{Flux}}$). 
    While PTMCMC consistently fails to converge (high JSD), FluxMC achieves high-fidelity recovery across the full parameter space.
    }
    \label{tab:js_comparison_full}
    \setlength{\tabcolsep}{0.95pt}   
    \renewcommand{\arraystretch}{1.1} 
    \small 
    
    \begin{tabular}{l ccc ccc}
    \toprule
    & \multicolumn{3}{c}{\textbf{Simulation Case III}} & \multicolumn{3}{c}{\textbf{Simulation Case IV}} \\
    \cmidrule(lr){2-4} \cmidrule(lr){5-7} 
    
    \textbf{Parameter} & \textbf{PTMCMC} & \textbf{FluxMC} & \textbf{Improv.} & \textbf{PTMCMC} & \textbf{FluxMC} & \textbf{Improv.} \\
    
    & (JSD) & (JSD) & (Factor) & (JSD) & (JSD) & (Factor) \\
    \midrule
    
    $\log_{10}\mathcal{M}_{c,z}$ (Chirp Mass) & 0.582 & 0.003 & $179\times$ & 0.485 & 0.004 & $136\times$ \\
    $q$ (Mass Ratio)                          & 0.641 & 0.001 & $590\times$ & 0.685 & 0.003 & $231\times$ \\
    $\chi_{z,1}$ (Primary Spin)               & 0.633 & 0.001 & $837\times$ & 0.664 & 0.003 & $231\times$ \\
    $\chi_{z,2}$ (Secondary Spin)             & 0.635 & 0.013 & $48\times$  & 0.690 & 0.002 & $300\times$ \\
    $t_c$ (Coalescence Time)                  & 0.840 & 0.046 & $18\times$  & 0.298 & 0.004 & $71\times$ \\
    $\varphi_c$ (Phase)                       & 0.616 & 0.005 & $122\times$ & 0.544 & 0.003 & $194\times$ \\
    $\log_{10} D_L$ (Distance)                & 0.827 & 0.036 & $23\times$  & 0.691 & 0.003 & $203\times$ \\
    $\cos \iota$ (Inclination)                & 0.681 & 0.026 & $26\times$  & 0.693 & 0.004 & $192\times$ \\
    $\lambda$ (Ecliptic Longitude)            & 0.807 & 0.016 & $50\times$  & 0.259 & 0.003 & $96\times$ \\
    $\sin \beta$ (Ecliptic Latitude)          & 0.911 & 0.038 & $24\times$  & 0.693 & 0.003 & $231\times$ \\
    $\psi$ (Polarization)                     & 0.815 & 0.028 & $29\times$  & 0.693 & 0.003 & $203\times$ \\
    \midrule
    
    \textbf{Average (All 11 Params)} & \textbf{0.726} & \textbf{0.019} & \textbf{$\sim 38\times$} & \textbf{0.581} & \textbf{0.003} & \textbf{$\sim 180\times$} \\
    \bottomrule
    \end{tabular}
\end{table*}

Collectively, these results remove a critical impasse in high-dimensional inference, demonstrating that FluxMC enables rapid and reliable high-fidelity science data analysis for all planned space-based gravitational-wave observatories. Beyond mere acceleration, the case studies presented in this chapter comprehensively validate FluxMC as an efficient, reliable, and practical solution for the most challenging complex parameter inversion problems. 
By effectively navigating the multimodal landscapes that entrap conventional algorithms, FluxMC ensures the robust inclusion of HMs in routine analysis. 
FluxMC is poised to unlock the full scientific potential of higher-order modes by breaking the fundamental parameter degeneracies that limit standard methods.
As evidenced by the substantial improvements in estimation precision achieved in our analyses, FluxMC enhances parameter recovery by 1-2 orders of magnitude, establishing a new paradigm where the scientific yield of future space-borne missions is limited only by instrument sensitivity, not by algorithmic constraints.

\section{Conclusion}

The analysis of simulated MBHB signals in this work exposes a fundamental tension in scientific computing: the conflict between the need for high-fidelity physical models and the inherent limitations of conventional inference algorithms. We demonstrate that traditional PTMCMC methods face a critical impasse when confronted with complex, high-dimensional landscapes—not only due to the prohibitive cost of likelihood evaluations but, more fundamentally, due to the vulnerability of local-walking samplers to become trapped in suboptimal modes.

In this work, we introduce FluxMC to address this pervasive dilemma. By leveraging Flow Matching to learn the complex multimodal landscape of the posterior, FluxMC shifts the Bayesian inference paradigm from a blind, local ``search'' into a globally guided transport. This framework comprehensively validates FluxMC as an efficient and reliable solution for challenging parameter inversion problems. Its ability to navigate multi-modal landscapes without prior knowledge of the mode locations ensures robust convergence where standard samplers fail, offering a practical pathway to overcome local-optima entrapment in high-dimensional Bayesian analysis.

Consequently, FluxMC significantly reduces the need to compromise between model accuracy and analysis speed. By enabling full, unbiased inference with complex, high-fidelity models, it is poised to help unlock the scientific potential of future space-borne missions like LISA and Taiji. Beyond gravitational-wave astronomy, this globally guided exact inference framework may be applicable to other computationally expensive, multi-modal inverse problems. Future work will focus on validating this approach across a broader range of scientific domains and rigorously testing it against real observational data.

\section{Methodological Framework}\label{Sec:model}

\subsection{Frequency-domain GW Signal Response and Instrumental Noise Model}

\begin{figure}[htb!]
    \centering
    \includegraphics[width=0.95\textwidth]{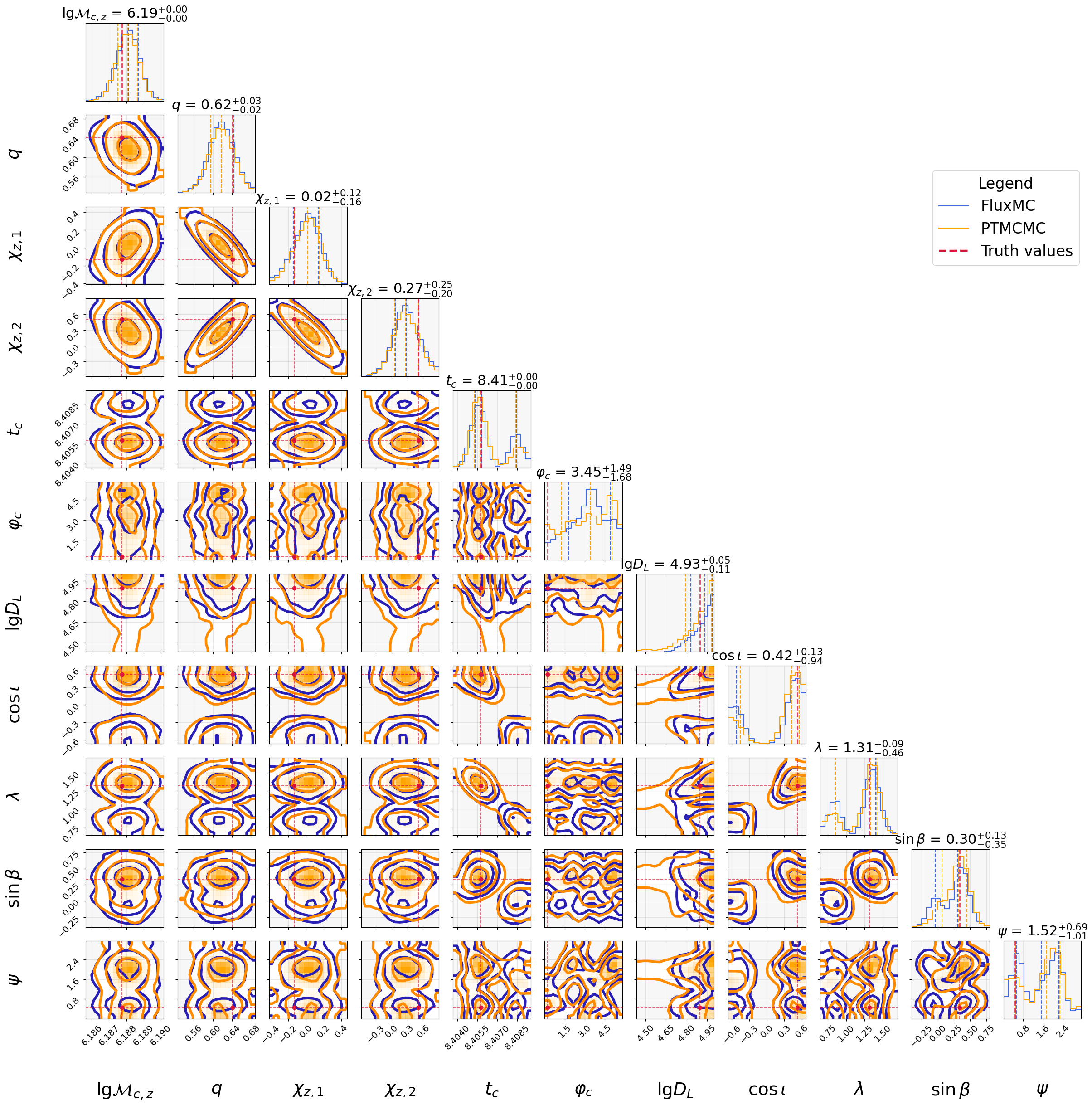}
    \caption{
        \textbf{Complete Posterior Distributions for Accelerated, Unbiased Inference.}
        This figure shows the full parameter corner plot corresponding to the analysis in Fig.~\ref{fig:lisa_unbiased_d_model}. It compares the parameter estimation for a simulated MBHB signal, using the computationally efficient \texttt{IMRPhenomD} template.
        Both \textbf{FluxMC} (blue contours and histograms) and the conventional PTMCMC (orange contours and histograms) successfully recover the true injected parameters (red dashed lines). The results are statistically identical, confirming FluxMC's accuracy.
    }
    \label{fig:lisa_unbiased_d_full}
\end{figure}

\begin{figure}[htb!]
    \centering
    \includegraphics[width=0.9\textwidth]{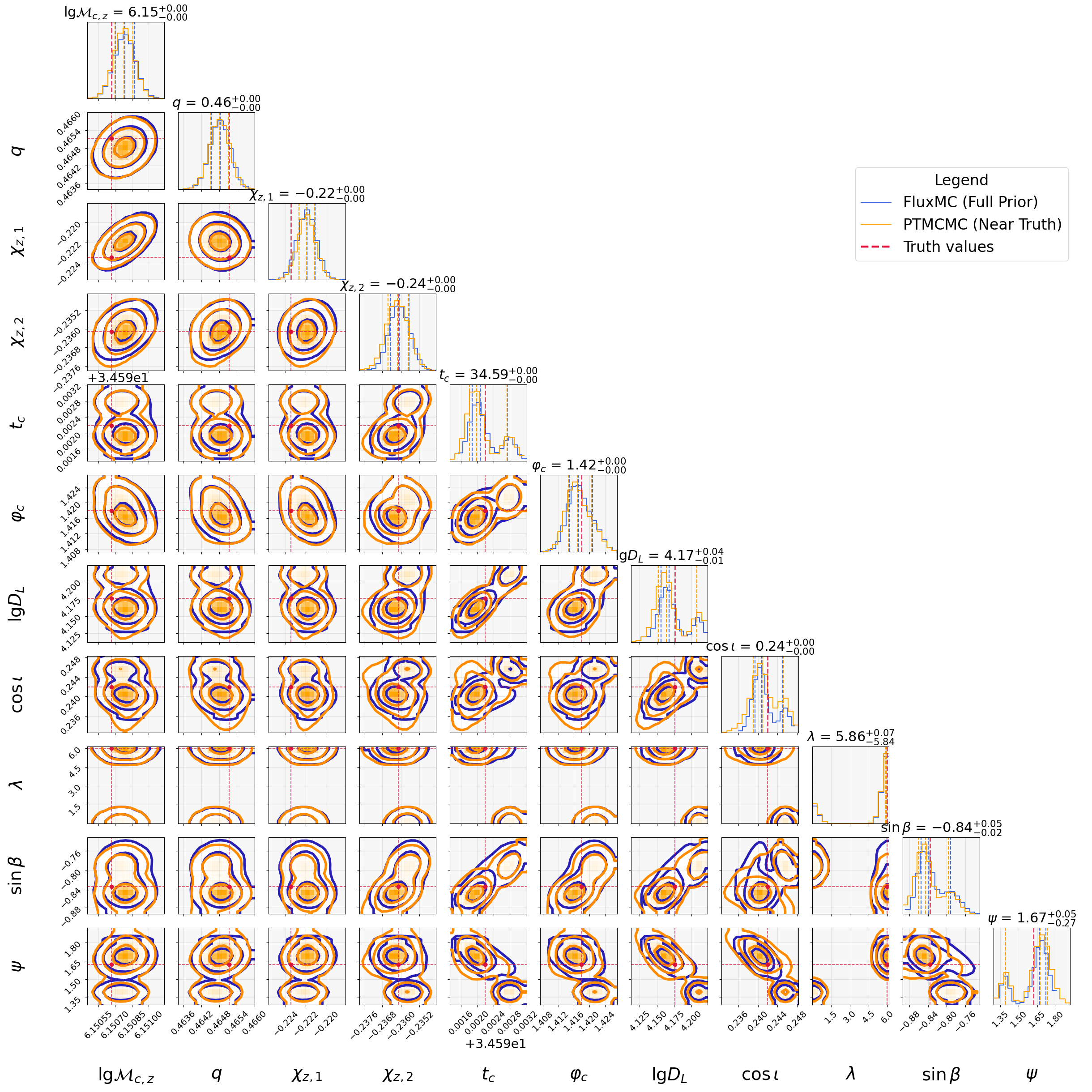}
    \caption{
        \textbf{Complete Posterior Distributions for \texttt{IMRPhenomHM} Validation.}
        This figure shows the full 11-parameter corner plot corresponding to the validation analysis in Fig.~\ref{fig:hm_comparison}(a).
        The results from \textbf{FluxMC (Full Prior)} (blue contours and histograms) are shown to be statistically identical to the \textbf{PTMCMC (Near Truth)} benchmark (orange contours and histograms).
        Both methods successfully recover the true injected parameters (red dashed lines), confirming FluxMC's accuracy for high-fidelity \texttt{IMRPhenomHM} inference on Case \textrm{III}.
    }
    \label{fig:lisa_hm_validation_full} 
\end{figure}

The \texttt{IMRPhenom} waveform family consists of various frequency-domain templates,  thus our analysis operates in the frequency domain accordingly.  
The data can be expressed as the superposition of signal  and noise, where the  model of GW signal involves two components: the GW  waveform and the detector response.
For the purpose of noise suppression, in space-based GW detection, the data used for scientific analysis are the so-called Time-Delay Interferometry (TDI) data streams, which are synthesized by applying appropriate time delays and linear combinations to the raw measurements~\cite{PhysRevD.71.022001,PhysRevD.69.082001, PhysRevD.62.042002}.
Considering the most commonly adopted 2nd-generation Michelson-$A_2, E_2$ TDI channels, the frequency-domain GW signal response can be formulated as~\cite{Yuan:2025wyx} 
\begin{equation}
    \widetilde{\rm TDI}(f) = \sum_{\ell m} \mathcal{G}^{\ell m}_{\rm TDI}(f, t_{f, \ell m}) \mathcal{A}_{\ell m}(f) e^{-i \Phi_{\ell m}(f)}, \quad {\rm TDI} \in \{A_2, E_2\}, 
    \label{eq:TDI_X2}
\end{equation}
where $\mathcal{A}_{\ell m}(f) e^{-i \Phi_{\ell m}(f)}$ is the waveform of  $(\ell, m)$ mode, with $\mathcal{A}_{\ell m}(f)$ and $\Phi_{\ell m}(f)$ being the amplitude and phase, respectively. 
The \texttt{IMRPhenomD} template includes only the primary $(\ell, m)=(2,2)$ mode~\cite{Husa:2015iqa,PhysRevD.93.044007}, while the $\texttt{IMRPhenomHM}$ template we employ accounts for 6 modes $(\ell, m) \in \{(2,2), (3,3), (4,4), (2, 1), (3, 2), (4, 3)\}$~\cite{Kalaghatgi:2019log,PhysRevLett.120.161102}. 
The increase in computational cost for \texttt{IMRPhenomHM} compared to \texttt{IMRPhenomD} is basically proportional to these additional mode numbers.
The  transfer function $\mathcal{G}^{\ell m}_{\rm TDI}(f, t_{f, \ell m})$ encapsulates the time-varying configuration of  detector caused by orbital dynamics,  and the time-frequency relationship for the $\ell m$ mode is  
\begin{equation}
t_{f, \ell m} \equiv  -\frac{1}{2\pi}\frac{d \Phi_{\ell m}}{df}. 
\end{equation}
Eq.~(\ref{eq:TDI_X2}) can be fully determined with 11 parameters: $\mathcal{M}_{c,z}$ is the redshifted chirp mass, $q$ is the mass ratio of binary, $\chi_{z1}, \chi_{z_2}$ represent the spins of two massive black holes, $t_c$ and $\varphi_c$ are the time and orbit phase at coalescence, $\iota$ denotes the inclination angle, $\psi$ is the polarization angle, and the sky location of MBHB is specified by $\{D_L, \lambda, \beta\}$, with $D_L$ being the luminosity distance, and $\lambda, \beta$ the Ecliptic longitude and latitude, respectively. 
Ref.~\cite{marsat2021exploring} provides a detailed theoretical analysis on the multimodality in the MBHB  parameter space. 
For angular position $\{\lambda, \beta\}$, 
when accounting for the detector motion and full response model (\emph{i.e.} no low-frequency approximation), depending on the relative orientation of detector and source,   a ``reflected''  mode can appear (see \emph{e.g.} Fig.~\ref{fig:lisa_unbiased_d_full}). 
While for the phase parameter $\varphi_c$, 
due to the distinct manner in which $\varphi_c$ enters the response for different modes, the  multimodality observed in \texttt{IMRPhenomD} waveform is effectively resolved once  HMs are included (\emph{e.g.} comparing Fig.~\ref{fig:lisa_unbiased_d_full} with  Fig.~\ref{fig:lisa_hm_validation_full}).


The noise budgets of both LISA and Taiji comprise two primary components: the Optical Metrology System (OMS) noise and the test-mass residual ACCeleration (ACC) noise. 
Both mission shares the same noise spectral model, while their respective noise budget parameters are different.
For LISA~\cite{Colpi2024:lisa}, the nominal arm length is $L = 2.5 \times 10^9 \ \mathrm{m}$ (corresponding to laser propagation time $d = 8.3 \ \mathrm{s}$), with noise amplitudes $A_{\rm OMS} = 15 \times 10^{-12} \ \mathrm{m}/\sqrt{\mathrm{Hz}}$ and $A_{\rm ACC} \approx 3 \times 10^{-15} \ \mathrm{m}/\mathrm{s}^2/\sqrt{\mathrm{Hz}}$.
For Taiji~\cite{10.1093/ptep/ptaa083}, the nominal arm length is $L = 3 \times 10^9 \ \mathrm{m}$ ($d \approx 10 \ \mathrm{s}$), with  OMS requirement of $A_{\rm OMS} = 8 \times 10^{-12} \ \mathrm{m}/\sqrt{\mathrm{Hz}}$ and  acceleration noise $A_{\rm ACC} = 3 \times 10^{-15} \ \mathrm{m}/\mathrm{s}^2/\sqrt{\mathrm{Hz}}$.
The respective Power Spectral Densities (PSDs) for these two noise  components  are given by
\begin{subequations}
\begin{align}
    P_{\rm OMS}(f) &= A_{\rm OMS}^2 \left(\frac{2\pi f}{c}\right)^2 \left[1 + \left(\frac{2 \ \mathrm{mHz}}{f}\right)^4\right], \label{eq:P_OMS} \\
    P_{\rm ACC}(f) &= A_{\rm ACC}^2 \left(\frac{1}{2\pi f c}\right)^2 \left[1 + \left(\frac{0.4 \ \mathrm{mHz}}{f}\right)^2\right] \left[1 + \left(\frac{f}{8 \ \mathrm{mHz}}\right)^4\right], \label{eq:P_ACC}
\end{align}
\end{subequations}
where the factors $(2\pi f /c )^2$ and $[1 / (2\pi f c)]^2$ convert displacement and acceleration units to fractional frequency shift units. 
The total noise in the analyzed data stream arises from the propagation of these component instrumental noises through TDI combinations. 
Assuming the noise sources are identical and uncorrelated among spacecrafts, under the equal-arm approximation, we derive the noise PSDs for the TDI-$A_2, E_2$ channels as
, they read
\begin{equation}
\begin{split}
    P_{A_2}(f) \approx P_{E_2}(f) \approx & \ 32\sin^2 (2u) \sin^2(u) \bigg[\left(2 + \cos(u)\right) P_{\rm OMS}(f) \\
    & + \left(6 + 4 \cos(u) + 2\cos(2u) \right) P_{\rm ACC}(f) \bigg],
\end{split}
\label{eq:P_A2E2}
\end{equation}
where $u \equiv 2\pi f d$ is the dimensionless frequency normalized by the laser propagation time $d$.

\subsection{On-the-fly Generation and Adaptive Noise Strategy}

To ensure the neural network adapts to signal characteristics across the full parameter space and achieves high robustness against instrumental noise, we implement an adaptive noise strategy that integrates on-the-fly waveform generation, frequency-domain whitening, and dynamic noise injection. 
Unlike methods relying on fixed offline datasets, we utilize an ``infinite training horizon'' approach where a fresh batch of waveforms $\tilde{d}_{\mathrm{signal}}$ is synthesized via GPU acceleration at each training step~\cite{Du:2025xdq}. The source parameters $\theta$ for these waveforms are randomly sampled from the prior distributions detailed in Table~\ref{table:priors}, ensuring the model continuously encounters unique samples and learns the generalized mapping between physical parameters and waveforms.

\begin{table*}[htb]
    \footnotesize
    \caption{Prior distributions of the source parameters used in this work. 
    The table lists the lower and upper bounds for each parameter. 
    Note that while most parameters are sampled uniformly within the bounds, the chirp mass $\mathcal{M}_{c,z}$ and luminosity distance $d_L$ follow a log-uniform distribution, and the inclination $\iota$ and latitude $\beta$ are sampled isotropically (uniform in $\cos\iota$ and $\sin\beta$, respectively).}
    \label{table:priors}
    \setlength{\tabcolsep}{0pt} 
    \begin{tabular*}{\textwidth}{@{\extracolsep{\fill}}clccc}
    \toprule
        \hline
        \textbf{Parameter} & \textbf{Description} & \textbf{Prior Lower Bound} & \textbf{Prior Upper Bound}& \textbf{Units} \\
        \hline
        $\mathcal{M}_{c,z}$ & Redshifted chirp mass & $10^{5.5}$ & $10^{6.5}$ & $M_\odot$ \\
        $q$ & Mass ratio ($m_2/m_1$) & $0.1$ & $0.9$ & -- \\
        $\chi_{z1}$ & Spin of the heavier BH ($z$-axis) & $-0.9$ & $0.9$ & -- \\
        $\chi_{z2}$ & Spin of the lighter BH ($z$-axis) & $-0.9$ & $0.9$ & -- \\
        $t_c$ & Time of coalescence & $t_{\mathrm{GPS}} - 0.01$ & $t_{\mathrm{GPS}} + 0.01$ & day \\
        $\varphi_c$ & Phase at coalescence & $0$ & $2\pi$ & rad \\
        $D_L$ & Luminosity distance & $10^4$ & $10^5$ & Mpc \\
        $\iota$ & Inclination angle & $0$ & $\pi$ & rad \\
        $\lambda$ & Ecliptic longitude & $0$ & $2\pi$ & rad \\
        $\beta$ & Ecliptic latitude & $-\pi/2$ & $\pi/2$ & rad \\
        $\psi$ & Polarization angle & $0$ & $\pi$ & rad \\
        \hline
    \bottomrule
    \end{tabular*}
\end{table*}

Following generation, the signals undergo frequency-domain whitening to standardize amplitudes and mitigate colored noise effects. Using the theoretical noise power spectral density $S_n(f)$, the whitened signal is defined as:
\begin{equation}
    \tilde{d}_{\mathrm{white}}(f) = \frac{\tilde{d}_{\mathrm{signal}}(f)}{\sqrt{S_n(f)} \times \kappa},
    \label{eq:whitening}
\end{equation}
where $\kappa = \sqrt{T_{\mathrm{obs}}/4}$ is the normalization factor. This transformation effectively maps the instrumental noise distribution to a standard complex Gaussian with zero mean and unit variance.

Finally, to simulate the stochastic nature of real detector noise, we employ dynamic adaptive noise injection. For each training sample, independent random noise realizations $\tilde{n}_{\mathrm{real}}, \tilde{n}_{\mathrm{imag}} \sim \mathcal{N}(0, 1)$ are superimposed onto the whitened signals:
\begin{equation}
    \tilde{d}_{\mathrm{input}} = \tilde{d}_{\mathrm{white}} + (\tilde{n}_{\mathrm{real}} + \mathrm{i} \cdot \tilde{n}_{\mathrm{imag}}).
    \label{eq:noise_injection}
\end{equation}
This ensures that even identical physical waveforms $\tilde{h}(\theta)$ are paired with different noise realizations across epochs, enabling the model to distinguish intrinsic signal features from incidental noise fluctuations and enhancing its robustness for real observational data.

\subsection{Methodological Framework: Continuous Normalizing Flows and Flow Matching}
\label{sec:method_framework}

While current machine learning applications in natural sciences frequently employ Neural Posterior Estimation (NPE)~\cite{papamakarios2018fastepsilonfreeinferencesimulation, Dax_2021} with Discrete Normalizing Flows (DNFs), recent advances suggest that FM with Continuous Normalizing Flows (CNFs)~\cite{chen2019neuralordinarydifferentialequations}—termed Flow Matching Posterior Estimation (FMPE)~\cite{Dax2023FlowMF, lipman2022flow}—offers superior training efficiency and accuracy~\cite{Liang_2024, liang2024acceleratingstochasticgravitationalwave}.

The fundamental objective of CNFs is to construct a diffeomorphism $f: \mathbb{R}^d \to \mathbb{R}^d$ that transports a simple base distribution $q_0(\bm{\theta})$ (e.g., a standard normal distribution) to a complex target posterior $q_1(\bm{\theta}|\bm{d})$, where $\bm{d}$ represents the observational data. CNFs realize this transformation via a continuous process governed by a temporal parameter $t \in [0,1]$, where the dynamics are defined by a learnable neural vector field $v_{\phi}(t, \bm{\theta}, \bm{d})$. The trajectory of a sample $\bm{\theta}_t$ follows the Ordinary Differential Equation (ODE):
\begin{equation}
    \frac{\mathrm{d}\bm{\theta}_t}{\mathrm{d}t} = v_{\phi}(t, \bm{\theta}_t, \bm{d}), \quad \bm{\theta}_0 \sim q_0(\bm{\theta}), \quad \bm{\theta}_1 \sim q(\bm{\theta}|\bm{d}).
    \label{eq:ode_dynamics}
\end{equation}
The associated probability density evolution follows the continuity equation, enabling instantaneous density evaluation:
\begin{equation}
    q(\bm{\theta}_1|\bm{d}) = q_0(\bm{\theta}_0) \exp\left(-\int_0^1 \nabla_{\bm{\theta}} \cdot v_{\phi}(t, \bm{\theta}_t, \bm{d}) \mathrm{d}t\right).
    \label{eq:density_evolution}
\end{equation}

To train this continuous flow efficiently, we adopt the Conditional FM paradigm. We define a Linear Optimal Transport (OT) path for the intermediate state $\bm{\theta}_t$ between the source noise $\bm{\theta}_0$ and the ground-truth parameter $\bm{\theta}_1$:
\begin{equation}
    \bm{\theta}_t = (1 - t)\bm{\theta}_0 + t \bm{\theta}_1, \quad t \in [0, 1].
    \label{eq:ot_path}
\end{equation}
The target vector field generating this path is constant: $u_t(\bm{\theta}_t|\bm{\theta}_1) = \bm{\theta}_1 - \bm{\theta}_0$. Consequently, the training objective (Loss Function) of FluxMC is to minimize the expected mean squared error between the network output and this target field:
\begin{equation}
    \mathcal{L}_{\mathrm{FM}}(\phi) = \mathbb{E}_{t, q_0(\bm{\theta}_0), q(\bm{\theta}_1)} \left[ \left\| v_{\phi}(t, \bm{\theta}_t, \bm{d}) - (\bm{\theta}_1 - \bm{\theta}_0) \right\|^2 \right].
    \label{eq:fmpe_loss}
\end{equation}
By minimizing this loss, the network learns to regress straight-line trajectories from noise to data, resulting in highly stable and fast numerical integration during inference.

\subsection{FluxMC Network Architecture}
\label{sec:network_arch}

The time-dependent vector field $v_{\phi}(t, \bm{\theta}_t, \bm{d})$ is parameterized by a custom deep neural network designed to capture the complex conditional dependencies between the observational data and physical parameters. The architecture integrates data conditioning, temporal embedding, and a deep regression backbone into a unified framework.

First, to handle the high-dimensional frequency-domain strain data $\tilde{d}_{\mathrm{input}}$, we employ a conditioner network composed of a sequence of Dense Residual Blocks. This module functions as a feature extractor, compressing the raw waveform into a compact, informative latent context vector $\bm{c}$ that encapsulates the essential physical properties of the signal. Simultaneously, the continuous time variable $t$ is mapped to a high-dimensional feature space using sinusoidal positional encoding, ensuring the network remains sensitive to the specific stage of the flow evolution.

These conditioning features—the data context $\bm{c}$ and the temporal embedding—are concatenated with the transient parameter state $\bm{\theta}_t$ to form the input for the primary regressor. For the backbone, we utilize a Residual Multi-Layer Perceptron (ResMLP). By leveraging deep residual connections, the ResMLP effectively mitigates gradient vanishing issues during training and accurately outputs the velocity field required to guide the optimal transport trajectory from the prior to the posterior.

\subsection{FluxMC Inference Strategy}
\label{sec:inference_strategy}

To harmonize real-time inference speed with rigorous, unbiased Bayesian statistics, we propose the FluxMC strategy, which conceptually utilizes the flow model as a ``global navigator'' and MCMC as a ``local corrector.'' 
The inference process commences with a flow-based fast global search. Upon detecting gravitational wave data $\bm{d}_{\mathrm{obs}}$, the pre-trained FMPE network integrates the learned ODE (Eq.~\ref{eq:ode_dynamics}) to transform random noise from a standard normal distribution into physical parameter samples $\bm{\theta}_{\mathrm{flow}}$ within seconds. These samples rapidly identify and cover the primary modes of the posterior, providing a high-fidelity global approximation.

Subsequently, we leverage this flow-generated posterior approximation to initialize the entire population of the PTMCMC. The initialization volume is explicitly determined by the ensemble configuration, comprising $N_T$ temperature ladders and $N_W$ walkers per ladder. We extract the first $N_{\mathrm{total}} = N_T \times N_W$ samples from the flow-generated posterior set $\bm{\theta}_{\mathrm{flow}}$ and assign them to the corresponding chains: $\bm{\theta}_{\mathrm{walker}}^{(i, j)} \leftarrow \bm{\theta}_{\mathrm{flow}}[k]$, where $k = 1, \dots, N_{\mathrm{total}}$. By seeding the full ensemble with these high-probability samples, the sampler bypasses the traditionally expensive burn-in phase and immediately enters an effective sampling state.

Ultimately, scientific rigor is guaranteed through exact likelihood correction. The MCMC sampler evolves the chains based on the exact physical likelihood function $\mathcal{L}(\bm{d}|\bm{\theta})$, effectively treating the neural network output as a high-quality proposal distribution. Any minor biases in the network predictions are rectified via the Metropolis-Hastings acceptance criterion:
\begin{equation}
    \alpha = \min \left( 1, \frac{\mathcal{L}(\bm{d}|\bm{\theta}') \pi(\bm{\theta}')}{\mathcal{L}(\bm{d}|\bm{\theta}) \pi(\bm{\theta})} \right).
    \label{eq:mcmc_alpha}
\end{equation}
This mechanism ensures that the final posterior distribution is asymptotically unbiased, adhering strictly to physical laws while retaining the computational efficiency of deep learning.

\bmhead{Acknowledgements} 
This study is supported by the National Key Research and Development Program of China (Grant No. 2021YFC2201901, Grant No. 2021YFC2203004, Grant No. 2020YFC2200100 and Grant No. 2021YFC2201903). 
International Partnership Program of the Chinese Academy of Sciences, Grant No. 025GJHZ2023106GC.
We also gratefully acknowledge the financial support from Brazilian agencies 
Funda\c{c}\~ao de Amparo \`a Pesquisa do Estado de S\~ao Paulo (FAPESP), 
Fundação de Amparo à Pesquisa do Estado do Rio Grande do Sul (FAPERGS),
Funda\c{c}\~ao de Amparo \`a Pesquisa do Estado do Rio de Janeiro (FAPERJ), 
Conselho Nacional de Desenvolvimento Cient\'{\i}fico e Tecnol\'ogico (CNPq), 
and Coordena\c{c}\~ao de Aperfei\c{c}oamento de Pessoal de N\'ivel Superior (CAPES). This work is also supported by High-performance Computing Platform of Peking University.

\bibliography{ref}

\section*{Appendix}
This appendix presents the complete posterior distribution plots (corner plots) corresponding to the analyses discussed in the main text. We provide detailed comparisons between FluxMC and the conventional PTMCMC sampler for simulated space-based GW data under different observatory configurations (Cases \textrm{I--IV}). The figures cover parameter estimation results using both IMRPhenomD and IMRPhenomHM waveform templates, illustrating the validation of unbiased inference and robustness tests under full prior ranges.

\begin{figure}[htb!]
    \centering
    \includegraphics[width=0.95\textwidth]{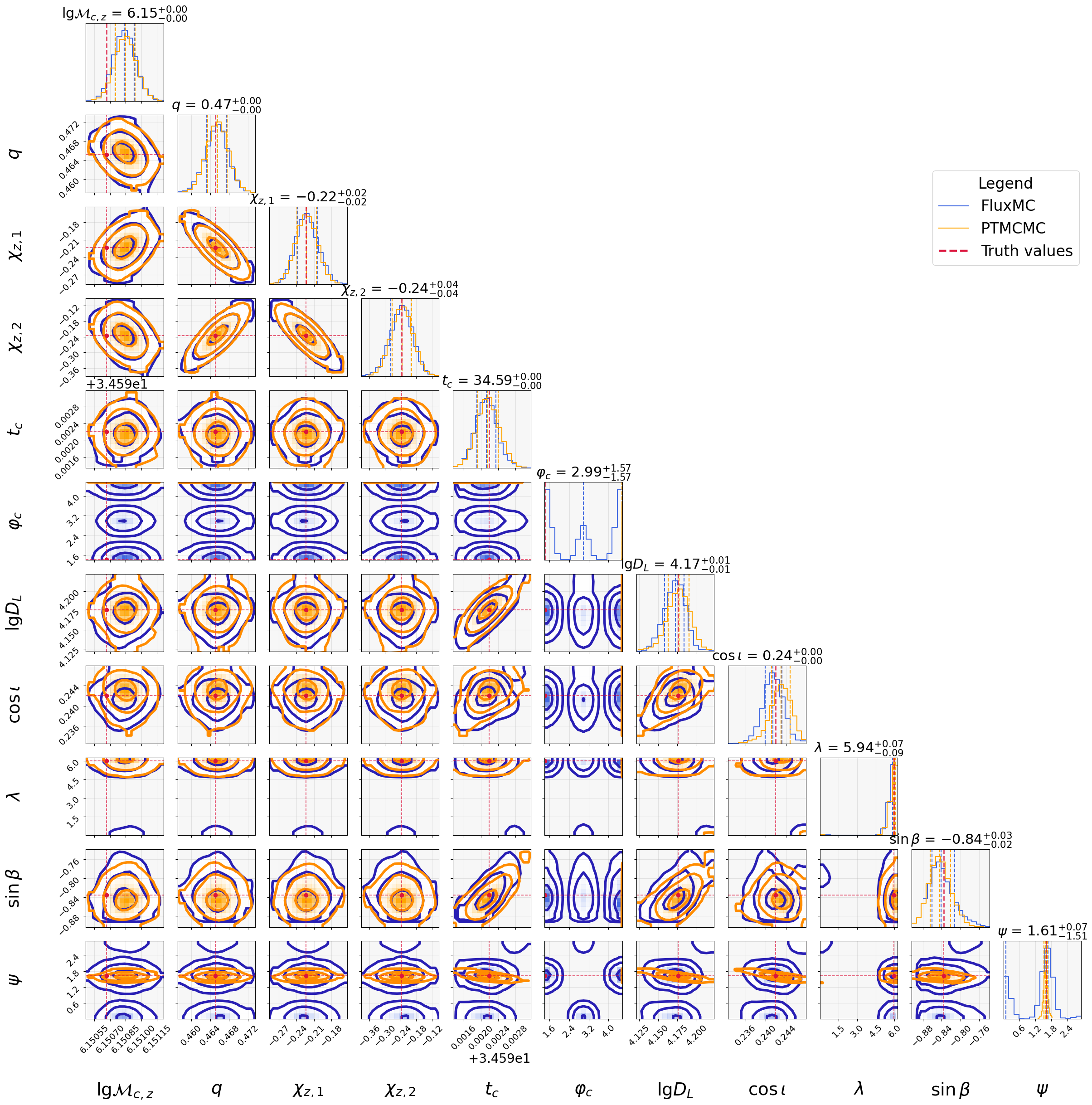}
    \caption{
        This figure shows the full parameter corner plot corresponding to the analysis in Fig.~\ref{fig:lisa_d_model}.
        The conventional PTMCMC sampler (orange contours and histograms) fails to find the true injected parameters (red dashed lines), producing a confidently biased result for extrinsic parameters like $\phi_c$.
        In contrast, \textbf{FluxMC} (blue contours and histograms) correctly explores the parameter space and recovers the true, unbiased posterior distributions.
    }
    \label{fig:lisa_biased_d_full}
\end{figure}

\begin{figure}[htb!]
    \centering
    \includegraphics[width=0.9\textwidth]{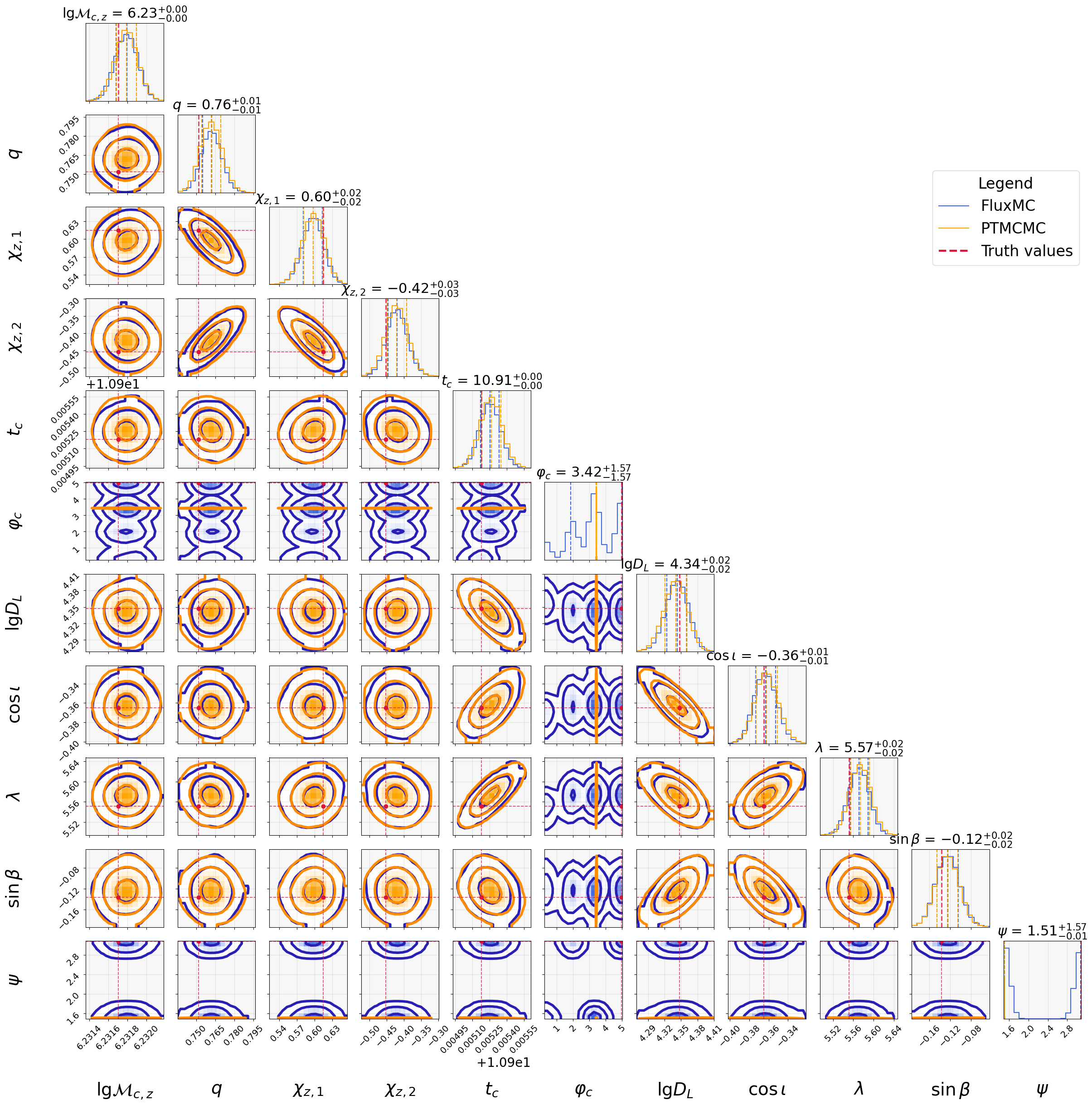}
    \caption{
        This figure shows the full parameter corner plot corresponding to the analysis in Fig.~\ref{fig:taiji_d_comparison}.
        The conventional PTMCMC sampler (orange contours and histograms) fails to find the true injected parameters (red dashed lines), producing a confidently biased result for extrinsic parameters like $\phi_c$ and $\psi$.
        In contrast, \textbf{FluxMC} (blue contours and histograms) correctly explores the parameter space and recovers the true, unbiased posterior distributions.
    }
    \label{fig:taiji_unbiased_d_full} %
\end{figure}

\begin{figure}[htb!]
    \centering
    \includegraphics[width=0.9\textwidth]{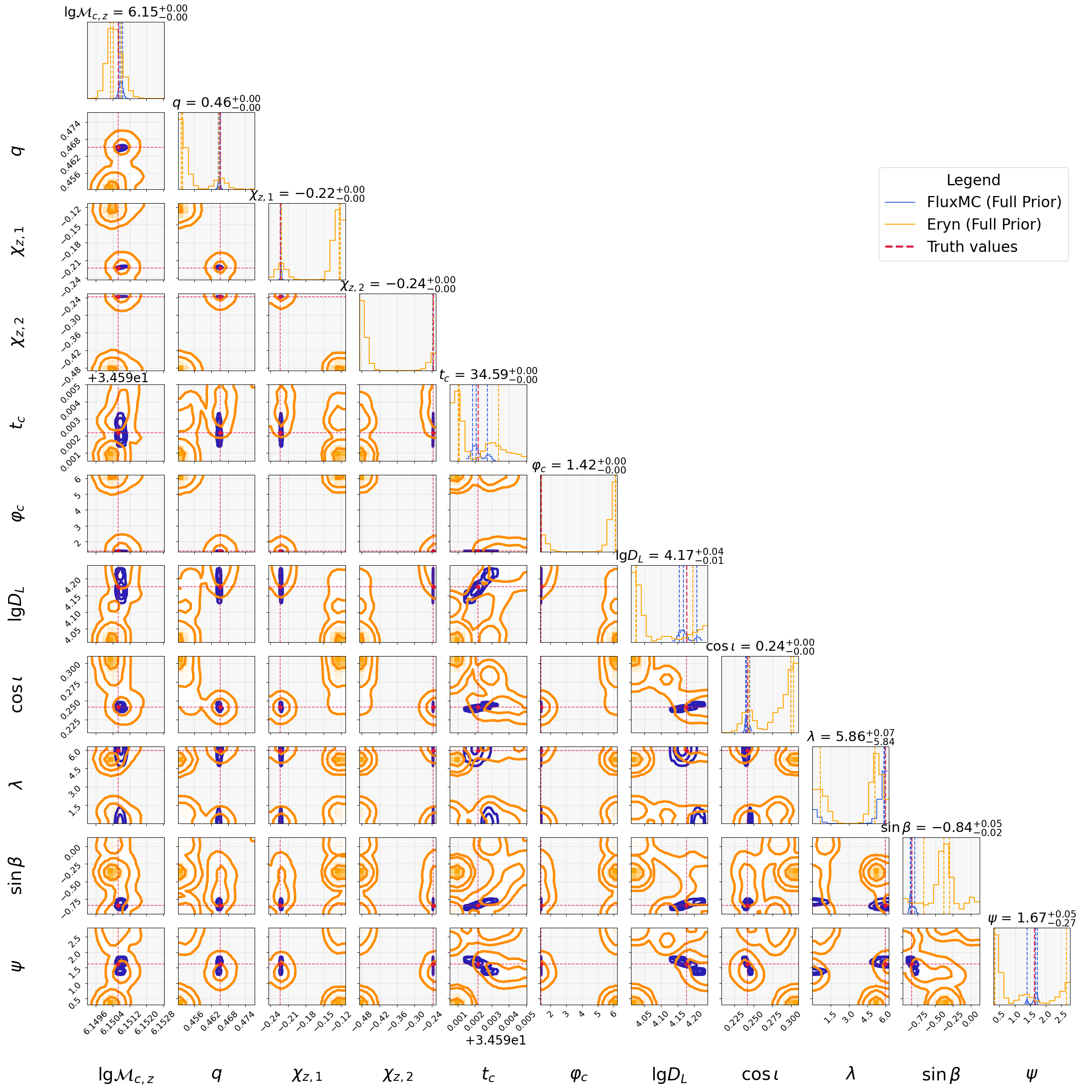}
    \caption{
        \textbf{Complete Posterior Distributions for \texttt{IMRPhenomHM} Robustness Test.}
        This figure shows the full 11-parameter corner plot corresponding to the robustness analysis in Fig.~\ref{fig:hm_comparison}(b).
        When starting from the full, unconstrained prior, \textbf{FluxMC (Full Prior)} (blue contours and histograms) successfully converges and recovers the true injected parameters (red dashed lines).
        In stark contrast, the conventional \textbf{PTMCMC (Full Prior)} sampler (orange contours and histograms) completely fails to converge, producing uninformative or biased posteriors that miss the true values for nearly all parameters.
    }
    \label{fig:lisa_hm_robustness_full} 
\end{figure}

\begin{figure}[htb!]
    \centering
    \includegraphics[width=0.9\textwidth]{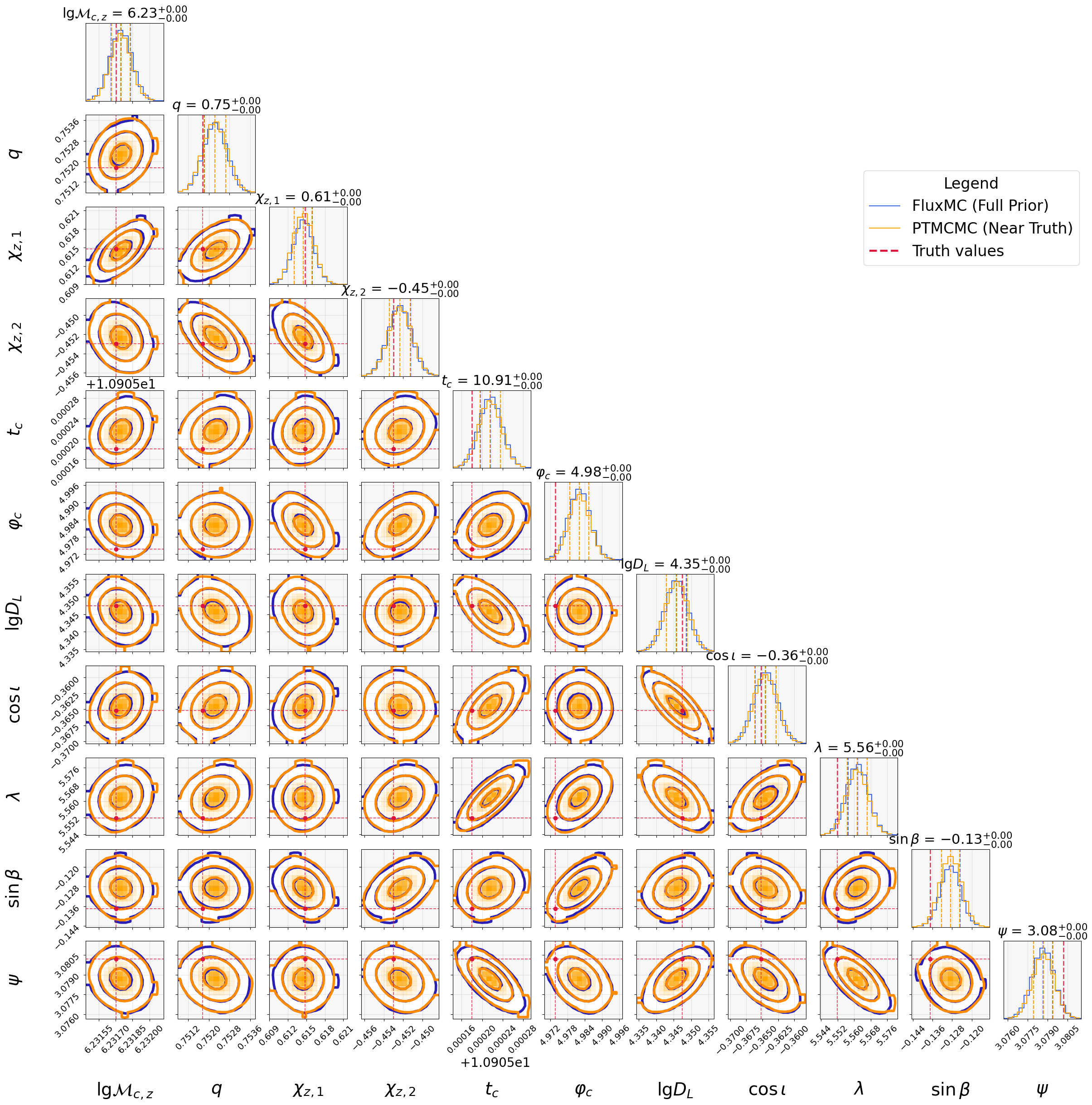}
    \caption{
        \textbf{Complete Posterior Distributions for \texttt{IMRPhenomHM} Validation.}
        This figure shows the full 11-parameter corner plot corresponding to the validation analysis in Fig.~\ref{fig:hm_comparison_taiji}(a).
        The results from \textbf{FluxMC (Full Prior)} (blue contours and histograms) are shown to be statistically identical to the \textbf{PTMCMC (Near Truth)} benchmark (orange contours and histograms).
        Both methods successfully recover the true injected parameters (red dashed lines), confirming FluxMC's accuracy for high-fidelity \texttt{IMRPhenomHM} inference on Case \textrm{IV}.
    }
    \label{fig:taiji_hm_validation_full} 
\end{figure}

\begin{figure}[htb!]
    \centering
    \includegraphics[width=0.9\textwidth]{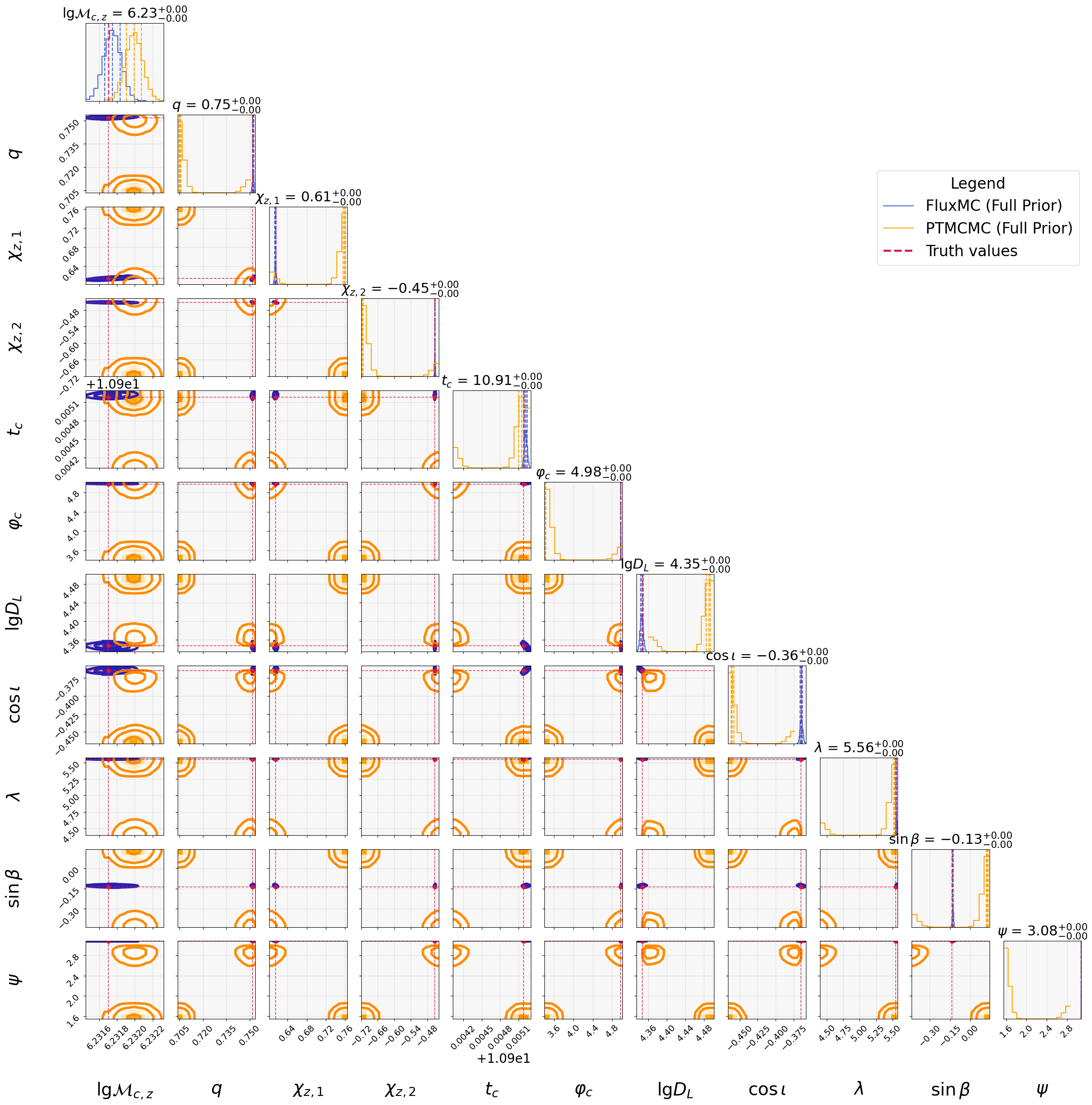}
    \caption{
        \textbf{Complete Posterior Distributions for \texttt{IMRPhenomHM} Robustness Test.}
        This figure shows the full 11-parameter corner plot corresponding to the robustness analysis in Fig.~\ref{fig:hm_comparison_taiji}(b).
        When starting from the full, unconstrained prior, \textbf{FluxMC (Full Prior)} (blue contours and histograms) successfully converges and recovers the true injected parameters (red dashed lines).
        In stark contrast, the conventional \textbf{PTMCMC (Full Prior)} sampler (orange contours and histograms) completely fails to converge, producing uninformative or biased posteriors that miss the true values for nearly all parameters.
    }
    \label{fig:taiji_hm_robustness_full} 
\end{figure}

\subsection{Posterior Calibration Test}
\label{sec:posterior_calibration}

To rigorously validate the statistical consistency and robustness of the FluxMC framework, we perform a posterior calibration test, commonly visualized via a Probability-Probability (P-P) plot.
We evaluate this by performing parameter estimation on a population of 40 simulated MBHB injections.
To cover diverse morphological features in the likelihood surface, the injection set is equally divided into 20 cases using the \texttt{IMRPhenomD} model and 20 cases using the \texttt{IMRPhenomHM} model, which incorporates higher-order multipoles.

For each injected event and each parameter, we calculate the percentile of the true injected value within the recovered one-dimensional marginalized posterior distribution. The cumulative distribution function (CDF) of these percentiles across all 40 events is then compared against the theoretical uniform distribution (represented by the diagonal line). We quantify this consistency using the Kolmogorov-Smirnov (KS) test. 

\begin{figure}[tbp]
    \centering
    \includegraphics[width=0.85\textwidth]{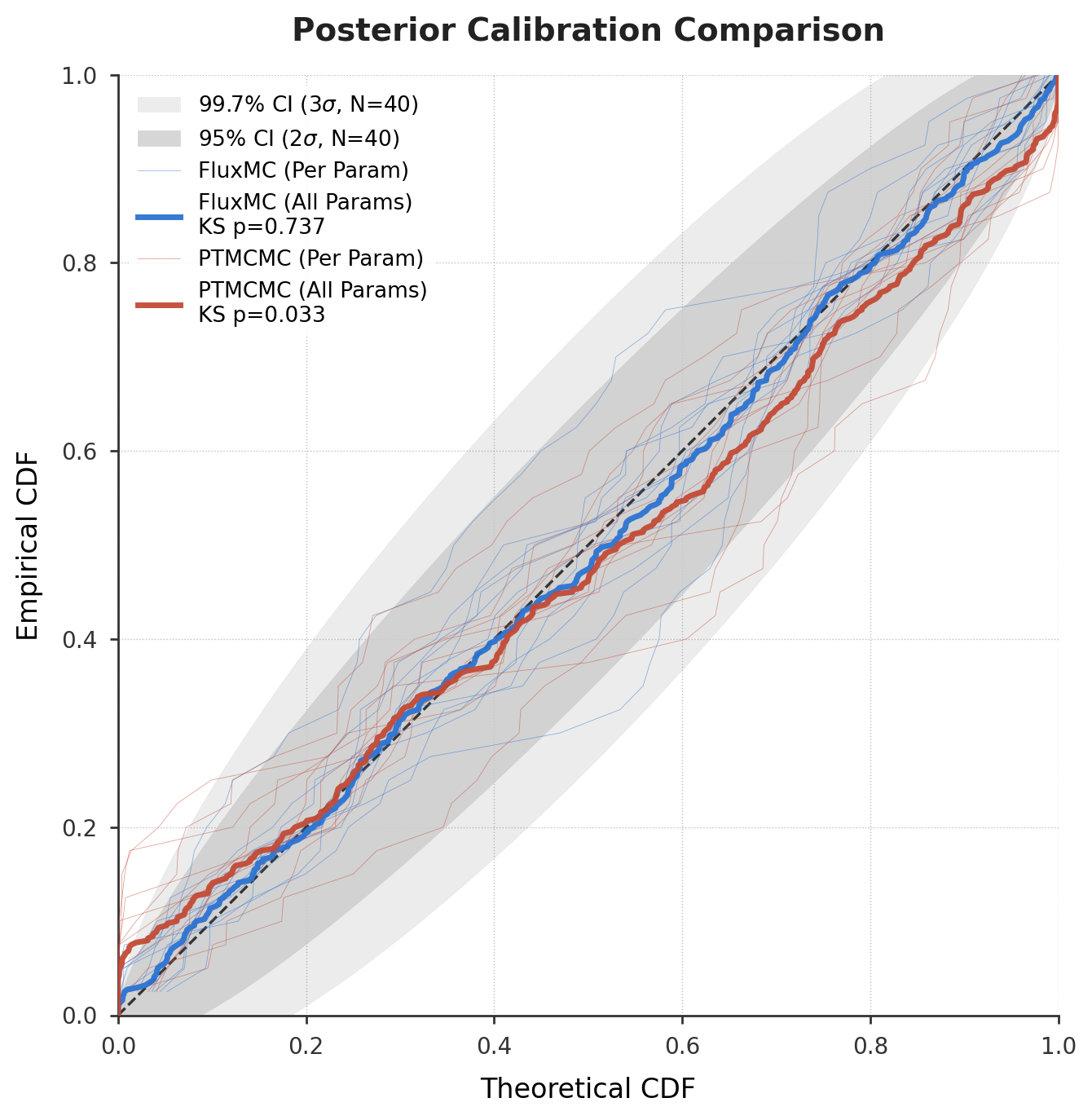} 
    \caption{\textbf{Posterior calibration (P-P) plot for 40 simulated MBHB injections across different waveform models.} 
    The injection set consists of 20 cases using the \texttt{IMRPhenomD} model and 20 cases using the \texttt{IMRPhenomHM} model. 
    The dashed diagonal line represents perfect statistical calibration, while the layered gray shaded regions indicate the expected $95\%$ ($\approx 2\sigma$) and $99.7\%$ ($\approx 3\sigma$) confidence intervals for $N=40$ independent trials. 
    Thick lines denote the aggregated results across all parameters, while thin lines show the results for individual parameters. 
    \textbf{FluxMC} (thick blue line) demonstrates near-perfect calibration ($p=0.737$), while the conventional \textbf{PTMCMC} sampler (thick red line) shows significant systematic bias ($p=0.033$), reflecting the impact of local-optima entrapment and incomplete mixing across the injected population.}
    \label{fig:pp_plot}
\end{figure}

As illustrated in Fig.~\ref{fig:pp_plot}, the aggregated FluxMC posteriors (thick blue line) demonstrate excellent calibration, tightly tracking the diagonal and yielding a high KS p-value of $p=0.737$. This indicates that FluxMC effectively navigates the multi-modal likelihood surfaces and breaks the parameter degeneracies introduced by higher-order modes in the \texttt{IMRPhenomHM} samples. In contrast, the conventional PTMCMC sampler (thick red line) exhibits significant systematic bias, with its CDF straying outside the $2\sigma$ confidence region and yielding a p-value of $p=0.033$. This miscalibration in PTMCMC is primarily driven by its susceptibility to local-optima entrapment and inefficient mixing when faced with increased topological complexity, further underscoring the necessity of the flow-guided global exploration provided by FluxMC.

\end{document}